\documentclass[a4paper,11pt]{article}
\pdfoutput=1
\usepackage{jheppub}
\usepackage{amsmath,amssymb,latexsym}
\usepackage{bm}
\usepackage{hepunits}
\usepackage[svgnames]{xcolor}

\usepackage{graphicx}
\usepackage{float}
\usepackage{subfigure}

\usepackage{mathtools}
\usepackage{braket}

\usepackage{natbib}
\usepackage{graphicx}

\usepackage{shuffle} 
\usepackage{slashed} 

\usepackage{cprotect}
\usepackage{multirow}
\usepackage{diagbox}

\newcommand{\zcal}{\mathcal{Z}}
\newcommand{\scal}{\mathcal{S}}

\DeclareUnicodeCharacter{2212}{-}
\DeclareUnicodeCharacter{2217}{*}

\newcommand{\nn}{\nonumber}

\newcommand{\mC}{\mathcal}

\def \lt {\left}
\def \rt {\right}
\def \eps {\epsilon}
\def \veps {\varepsilon}
\def \fc{\frac}
\def \td{\tilde}

\allowdisplaybreaks

\title{Two-loop QCD amplitudes for $t\bar{t}\gamma$ production at hadron colliders}

\author[a]{Guoxing Wang,}
\author[b]{Li Lin Yang}
\affiliation[a]{Laboratoire de Physique Th\'eorique et Hautes Energies (LPTHE), UMR 7589, Sorbonne Universit\'e et CNRS, 4 place Jussieu, 75252 Paris Cedex 05, France}
\affiliation[b]{Zhejiang Institute of Modern Physics, School of Physics, Zhejiang University, Hangzhou 310027, China}

\emailAdd{wangguoxing2015@pku.edu.cn}
\emailAdd{yanglilin@zju.edu.cn}

\abstract{The associated production of a photon and a top-antitop quark pair ($t\bar{t}\gamma$) is important for measuring the top-quark charge and probing the top-photon interaction, and it requires improved theoretical predictions. We focus on the calculation of two-loop amplitudes for $t\bar{t}\gamma$ production at hadron colliders. The infrared singularities with full top-quark mass dependence are derived from universal anomalous dimensions combined with one-loop massive amplitudes expanded to higher orders in the dimensional regulator $\epsilon$. The finite remainders are approximated in the high-energy boosted limit using the mass-factorization formula. To validate our approach, we compare approximate one-loop amplitudes up to $\mathcal{O}\left(\epsilon^2\right)$, as well as the two-loop infrared poles, against our exact results. The results in this paper serve as an important step toward next-to-next-to-leading order predictions for $t\bar{t}\gamma$ production.}

\begin{document}

\maketitle

\clearpage

\section{Introduction}

The associated production of a photon with a top-antitop quark pair ($t\bar{t}\gamma$) is one of the important processes at hadron colliders and next-generation experimental facilities. Evidence for $t\bar{t}\gamma$ production was first observed in $p\bar{p}$ collisions at the Tevatron by the CDF Collaboration \cite{CDF:2011ccg}. Subsequently, in $pp$ collisions at the Large Hadron Collider (LHC), both the ATLAS and CMS Collaborations observed and studied $t\bar{t}\gamma$ production \cite{ATLAS:2015jos, CMS:2017tzb, ATLAS:2017yax, ATLAS:2018sos, ATLAS:2020yrp, CMS:2021klw, CMS:2022lmh, ATLAS:2024hmk}. The $t\bar{t}\gamma$ process probes the top-photon interaction and provides a direct way to measure the top-quark electric charge \cite{Baur:2001si, Baur:2004uw}. Any possible modification of the top-photon vertex due to new physics would directly affect the production of $t\bar{t}\gamma$ (see, e.g., Refs.~\cite{Bouzas:2012av,BessidskaiaBylund:2016jvp,Schulze:2016qas,Etesami:2016rwu}), making precise measurements of this interaction an additional test of the Standard Model (SM). Furthermore, $t\bar{t}\gamma$ process also plays an important role in investigating the top-quark charge asymmetry \cite{Bergner:2018lgm,ATLAS:2022wec}.

Theoretical predictions for $t\bar{t}\gamma$ production at next-to-leading order (NLO) in quantum chromodynamics (QCD) with stable top quarks were first presented in Refs.~\cite{Duan:2009kcr,Duan:2011zxb,Maltoni:2015ena}, and later studied including top quark decays in Ref.~\cite{Melnikov:2011ta}. Matching to parton showers was addressed in Ref.~\cite{Kardos:2014zba}. Off-shell effects of the top quarks in this process were investigated in Refs.~\cite{Bevilacqua:2018woc,Bevilacqua:2018dny,Bevilacqua:2019quz,Stremmer:2024zhd}. NLO electroweak (EW) corrections were reported in Ref.~\cite{Duan:2016qlc}. Subsequently, the so-called complete NLO corrections with stable top quarks were presented in an automated calculation in Ref.~\cite{Pagani:2021iwa}, and later extended to include top quark decay in Ref.~\cite{Stremmer:2024ecl}. More recently, soft gluon resummation was performed and used to derive the approximate next-to-next-to-leading order (NNLO) predictions including both NLO QCD and NLO EW corrections~\cite{Kidonakis:2022qvz}. Beyond these developments, to match the unprecedented precision expected at current and future LHC runs, the full NNLO QCD corrections are essential and represent the current frontier. In these NNLO corrections, besides the two-loop amplitudes, higher order contributions from one-loop amplitudes in the dimensional regulator $\epsilon$ are also required. Recently, Ref.~\cite{Bera:2025upg} obtained the one-loop contributions up to $\mathcal{O}(\epsilon^2)$ for both the quark-antiquark annihilation and gluon fusion channels. In addition, a class of two-loop planar master integrals relevant for $t\bar{t}\gamma$ production was computed in Refs.~\cite{Badger:2022hno,Badger:2024fgb,Becchetti:2025oyb}\footnote{Although these works focus on $t\bar{t}j$ production, the master integrals considered therein also enter the $t\bar{t}\gamma$ process.}. Purely massless two-loop amplitudes have been considered in Ref.~\cite{Badger:2023mgf}. However, an exact calculation of the full two-loop amplitudes is still beyond the reach of current methods.

In this paper, we present approximate two-loop amplitudes for $t\bar{t}\gamma$ production. We first compute the infrared (IR) poles of the two-loop amplitudes using the method presented in Refs.~\cite{Ferroglia:2009ep, Ferroglia:2009ii}, which incorporates the full mass dependence of the top quark. For the finite terms, instead of performing an exact calculation, we work in the high-energy boosted limit where all kinematic invariants are much larger than the top-quark mass, i.e., $|s_{ij}|\gg m_t^2$ for all pairs of external legs $i\neq j$. In this regime, a general mass-factorization formula valid at leading power (LP) has been established in Refs.~\cite{Mitov:2006xs, Becher:2007cu, Engel:2018fsb, Wang:2023qbf}. To gain insight into power-suppressed effects—analogous to the study of $t\bar{t}H$ production in Ref.~\cite{Wang:2024pmv}—we keep part of the top-quark mass dependence in the amplitudes, leading to what we call ``semi-massive scheme''. Concretely, we start from the fully massive amplitudes but treat all scalar Feynman integrals as massless. In parallel, we construct amplitudes directly from LP mass factorization, which we refer to as the ``massless scheme''. These two approaches thus provide complementary approximate two-loop amplitudes for $t\bar{t}\gamma$ production. While significantly simpler than evaluating the fully massive amplitudes, both strategies are enabled by recent progress in integration-by-parts (IBP) reduction techniques \cite{Bohm:2018bdy, Bendle:2019csk, Boehm:2020ijp, Bendle:2021ueg, Wu:2023upw, Guan:2019bcx, Heller:2021qkz} and in the computation of two-loop five-point master integrals \cite{Gehrmann:2015bfy, Papadopoulos:2015jft, Gehrmann:2018yef, Chicherin:2018mue, Abreu:2018rcw, Abreu:2018aqd, Chicherin:2018old, Badger:2019djh}.

The paper is organized as follows. In Sec.~\ref{sec:convention}, we introduce the notation and define the quantities used in our calculation. In Sec.~\ref{sec:amp}, we derive the IR poles of the two-loop amplitudes with full mass dependence and present approximate formulas for the amplitudes in the high-energy boosted limit. Section~\ref{sec:squaredamp} discusses the treatment of power-suppressed contributions and the mapping between massive and massless phase-space points. Numerical results are shown in Sec.~\ref{sec:numeric}, and conclusions are given in Sec.~\ref{sec:conclusion}. Lengthy expressions are collected in the appendices.

\section{Notation and definitions}\label{sec:convention}
For photon production in association with a top-antitop quark pair, we consider the following two partonic processes:
\begin{align}
q_\beta(p_1)+\bar{q}_\alpha(p_2) &\to t_k(p_3)+\bar{t}_l(p_4)+\gamma(p_5) \,, \\
g_a(p_1)+g_b(p_2) &\to t_k(p_3)+\bar{t}_l(p_4)+\gamma(p_5) \,,
\end{align}
where $\alpha, \beta, a, b, k, l$ are color indices, and $p_i$ denote the momenta of the external partons with $p_1^2=p_2^2=p_5^2=0$ and $p_3^2=p_4^2=m_t^2$. The kinematic variables are defined as $s_{ij}=(p_i + \sigma_{ij} p_j)^2$, where $\sigma_{ij}=+1$ if $p_i$ and $p_j$ are both incoming or both outgoing, and $\sigma_{ij}=-1$ otherwise.

To facilitate the calculation of the amplitudes, we use the color-space formalism \cite{Catani:1996jh,Catani:1996vz}, in which the amplitudes are treated as vectors $\ket{{\cal M}_{q,g}}$. The subscripts $q$ or $g$ denote the quark-antiquark annihilation ($q\bar{q}$) channel or the gluon fusion ($gg$) channel, respectively. For the $q\bar{q}$ channel, we choose the independent color structures as
\begin{equation}\label{eq:qqbasis}
   \ket{c^q_1} = \delta_{\alpha\beta}\,\delta_{kl} \,, 
    \qquad 
   \ket{c^q_2} = (t^a)_{\alpha\beta}\,(t^a)_{kl} \,.
\end{equation}
For the $gg$ channel, we adopt the color basis
\begin{equation}\label{eq:ggbasis}
   \ket{c^g_1} = \delta^{ab}\,\delta_{kl} \,,
    \qquad 
   \ket{c^g_2} = if^{abc}\,(t^c)_{kl} \,,
    \qquad 
   \ket{c^g_3} = d^{abc}\,(t^c)_{kl} \,.
\end{equation}
Note that $\ket{c^q_I}$ and $\ket{c^g_I}$ are orthogonal but not normalized.
The amplitudes $\ket{\mC{M}_{q,g}}$ can therefore be decomposed in color space as
\begin{align}
\ket{\mC{M}_{q,g}} = \sum_{I}d_I^{q,g} \ket{c^{q,g}_I}\,,
\end{align}
where $d_I^{q,g}$ are the decomposition coefficients of the $I$-th color structure in the $q\bar{q}$ or $gg$ channel. We define the matrix elements of the squared amplitudes in this color space as
\begin{align}\label{eq:matirxE0}
\left(\bm{H}_{q,g}\right)_{IJ}&= \frac{\braket{c_I|{\cal M}_{q,g}}}{\braket{c_I|c_I}}\frac{\braket{{\cal M}_{q,g}|c_J}}{\braket{c_J|c_J}} \nonumber \\
&= d_{I,q,g}\left(d_{J,q,g}\right)^{\dag} \,,
\end{align}
where the spin sum over external partons has been performed with the polarization sums for the two initial gluons and the final photon chosen as
\begin{align}
\sum_s\veps^\mu(p_1,s)\veps^\nu(p_1,s)&=\sum_s\veps^\mu(p_2,s)\veps^\nu(p_2,s)=-g^{\mu\nu}+\frac{p_1^\mu p_2^\nu+p_1^\nu p_2^\mu}{p_1\cdot p_2} \,, \nonumber \\
\sum_s\veps^\mu(p_5,s)\veps^\nu(p_5,s)&=-g^{\mu\nu} \,.
\end{align}
The squared amplitudes in terms of $\left(\bm{H}_{q,g}\right)_{IJ}$ are then given by
\begin{align}\label{eq:sqamp0}
\braket{{\cal M}_{q,g}|{\cal M}_{q,g}}=\sum_{IJ}\left(\bm{H}_{q,g}\right)_{IJ}\braket{c_J|c_I} \,.
\end{align}

The ultraviolet (UV) divergences in the bare amplitudes $\ket{\mC{M}_{q,g}}$ are removed by renormalization,
\begin{align}
\Ket{{\cal M}^R_{q,g}(\alpha_s, e, m_t, \mu, \epsilon)} = \left(\frac{\mu^2 e^{\gamma_E}}{4\pi}\right)^{-\epsilon} Z_{q,g} Z_Q \Ket{{\cal M}^{\text{bare}}_{q,g}(\alpha_s^0, e^0, m_t^0,\epsilon)} ,
\label{eq:UVren}
\end{align}
where $Z_g$, $Z_q$, and $Z_Q$ are the on-shell wave-function renormalization constants for gluons, light quarks, and the heavy (top) quark, respectively, and dependence on other kinematic variables is left implicit.
The top-quark mass is renormalized in the on-shell scheme, $m_t^0 = Z_m m_t$, while the strong coupling constant $\alpha_s$ is renormalized in the $\overline{\text{MS}}$ scheme with $n_f=n_l+n_h$ active flavors:
\begin{equation}\label{eq:renorcoupling}
\alpha_s^0 = \left( \frac{\mu^2 e^{\gamma_E}}{4\pi} \right)^{\epsilon} Z_{\alpha_s} \alpha_s \, .
\end{equation}
Since we restrict ourselves to QCD corrections, the quantum electrodynamics (QED) coupling does not require renormalization, and we simply set $e^0=e$. Explicit expressions for the renormalization constants up to NNLO are collected in Appendix~\ref{sec:renormalconstant}.

After UV renormalization, the remaining IR divergences can be subtracted by a multiplicative factor $\bm{Z}^{-1}(\epsilon, m_t, \mu)$, where the boldface symbol denotes an operator in color space. Explicitly, we have
\begin{align}
\label{eq:key}
\bm{Z}_{q,g}^{-1}(\alpha_s, m_t, \mu, \epsilon) \Ket{{\cal M}^R_{q,g}(\alpha_s, e, m_t, \mu, \epsilon)} = \text{finite} \,.
\end{align}
Both the UV-renormalized amplitudes and the IR subtraction factors can be expanded in powers of $\alpha_s$:
\begin{align}
   \Ket{{\cal M}_{q,g}^R} &= 4\pi\alpha_s Q_te  \left[ \Ket{{\cal M}_{q,g}^{(0)}} 
    + \frac{\alpha_s}{4\pi} \Ket{{\cal M}_{q,g}^{(1)}} 
    + \left( \frac{\alpha_s}{4\pi} \right)^2 \Ket{{\cal M}_{q,g}^{(2)}} 
    + \cdots \right] , \nonumber
    \\
    \bm{Z}_{q,g} &= \bm{1} + \frac{\alpha_s}{4\pi} \bm{Z}_{q,g}^{(1)} + \left( \frac{\alpha_s}{4\pi} \right)^2 \bm{Z}_{q,g}^{(2)} + \cdots \, ,
\end{align}
where $Q_t = 2/3$ is the electric charge of the top quark. The IR subtraction factors up to NNLO can be found, for example, in Ref.~\cite{Ferroglia:2009ii}, and read
\begin{align}
   \bm{Z}^{(1)}_{q,g} &=  \frac{\Gamma_0'^{q,g}}{4\epsilon^2}
    + \frac{\bm{\Gamma}^{q,g}_0}{2\epsilon} \,,\label{eq:ZIRfactor1}\\
   \bm{Z}^{(2)}_{q,g} &= \frac{(\Gamma_0'^{q,g})^2}{32\epsilon^4} 
    + \frac{\Gamma_0'^{q,g}}{8\epsilon^3} 
    \left( \bm{\Gamma}^{q,g}_0 - \frac32\,\beta^{n_l}_0 \right) 
    + \frac{\bm{\Gamma}^{q,g}_0}{8\epsilon^2} 
    \left( \bm{\Gamma}^{q,g}_0 -2\beta^{n_l}_0 \right) 
    + \frac{\Gamma_1'^{q,g}}{16\epsilon^2}
    + \frac{\bm{\Gamma}^{q,g}_1}{4\epsilon} \nonumber \\
   &\qquad - \frac{2T_F}{3}n_h\, 
    \Bigg[ \Gamma_0'^{q,g} \left( \frac{1}{2\epsilon^2}\,\ln\frac{\mu^2}{m_t^2}
    + \frac{1}{4\epsilon} \left[ \ln^2\!\frac{\mu^2}{m_t^2}     
    + \frac{\pi^2}{6} \right] \right)
    + \frac{\bm{\Gamma}_0^{q,g}}{\epsilon}\,\ln\frac{\mu^2}{m_t^2} \Bigg] \,, \label{eq:ZIRfactor2}
\end{align}
where the coefficients $\bm{\Gamma}_n$ and $\Gamma_n'$ are defined via the expansion 
\begin{align}
   \bm{\Gamma} &= \sum_{n\ge 0}\,\left( \frac{\alpha_s}{4\pi} \right)^{n+1} \, \bm{\Gamma}_n \,, \qquad \Gamma_n' = \frac{\partial\bm{\Gamma}_n}{\partial\ln\mu}\,.
\end{align}
For $t\bar{t}\gamma$ production, the anomalous-dimension matrices $\bm{\Gamma}^{q,g}$ up to $\alpha_s^2$ in the color bases of Eqs.~\eqref{eq:qqbasis} and~\eqref{eq:ggbasis} coincide with those of $t\bar{t}H$ process~\cite{Broggio:2015lya,Wang:2024pmv}. We emphasize that in Eq.~\eqref{eq:ZIRfactor2}, $\beta^{n_l}_0 = 11\,C_A/3-4\,T_F n_l/3$ is the $\beta$-function coefficient, with $n_l$ denoting the number of active light flavors. The dependence on $n_h$ is entirely contained in the last line of Eq.~\eqref{eq:ZIRfactor2}. 

With these ingredients, the IR singularities of the amplitudes can be extracted order by order in $\alpha_s$:
\begin{align}
\label{eq:IRamps}
   \Ket{{\cal M}_{q,g}^{(1)\,,R,\,\text{sing}}} 
   &= \bm{Z}_{q,g}^{(1)} \Ket{{\cal M}_{q,g}^{(0)R}} , \nonumber
   \\
   \Ket{{\cal M}_{q,g}^{(2)\,,R,\,\text{sing}}} 
   &= \left[ \bm{Z}_{q,g}^{(2)} - \left( \bm{Z}_{q,g}^{(1)R}\right)^2 \right]
    \Ket{{\cal M}_{q,g}^{(0)R}}
    + \left( \bm{Z}_{q,g}^{(1)} \Ket{{\cal M}_{q,g}^{(1)R}} \right)_{\text{poles}} \,.
\end{align}
In particular, predicting the IR poles at two-loop order requires computing the UV-renormalized one-loop amplitudes up to ${\cal O}(\epsilon^1)$.

Having established the structure of the IR poles of the amplitudes, we now turn to the main objects of study in this work: the two-loop renormalized amplitudes $\ket{\mC{M}^{(2)R}_{q,g}}$. Their singular parts follow directly from Eq.~\eqref{eq:IRamps}, but the finite remainders require evaluating genuinely two-loop five-point Feynman integrals involving six physical scales—a task that is currently beyond exact analytic methods. To make progress, we adopt the high-energy boosted limit and employ the mass-factorization approach~\cite{Mitov:2006xs} to obtain approximate expressions for the finite parts. There is, however, a special class of two-loop diagrams with two fermion loops that is comparatively simple: the corresponding integrals can be factorized into products of two one-loop integrals.\footnote{This is achieved by expressing the two-loop integrals as linear combinations of master integrals via IBP reduction. In this case, all relevant master integrals can be chosen as products of two one-loop integrals.} These contributions correspond to the $n_l^2$, $n_h^2$, and $n_ln_h$ terms in  the color decomposition of Eq.~\eqref{eq:color_dec}, which are collectively denoted as $n_f^2$ terms in the following. We compute these contributions exactly, thereby providing both a check on the IR structure of $\ket{\mC{M}^{(2)R}_{q,g}}$ inferred from Eq.~\eqref{eq:IRamps} and a useful benchmark for our approximate results in the high-energy boosted limit.

\section{Calculation}\label{sec:amp}

\subsection{Massive amplitudes at one loop}

As discussed in the previous section, predicting the IR poles at two loops requires the full massive one-loop amplitudes expanded up to $\mathcal{O}\left(\epsilon^1\right)$. We generate the amplitudes with full top-quark mass dependence using \texttt{FeynArts} \cite{Hahn:2000kx}, and manipulate the resulting expressions with \texttt{FeynCalc} \cite{Shtabovenko:2020gxv,Shtabovenko:2023idz}. The amplitudes then need to be expressed in terms of scalar integrals. Since our main interest lies in the IR poles of the interference between $\mathcal{M}^{(2)}$ and $\mathcal{M}^{(0)}$ via Eq.~\eqref{eq:IRamps}, we can directly multiply the one-loop amplitudes by the tree-level ones. This allows the Lorentz contractions and Dirac traces to be carried out straightforwardly, while still keeping track of the color information necessary for predicting the two-loop amplitudes in the high-energy boosted limit according to the mass-factorization formula Eq.~\eqref{eq:QCDfac1}. Alternatively, one may apply a complete set of projectors to the one-loop amplitudes and extract the coefficients as linear combinations of scalar integrals. Although this method is more involved for our purposes, it is useful for obtaining the one-loop squared amplitude. We have performed the calculation using both approaches, and found consistent results.

The one-loop scalar integrals can be categorized into $17$ families (topologies) neglecting permutations of the external momenta among $p_1$, $p_2$, and $p_5$, and/or between $p_3$ and $p_4$. Each family is specified by five propagators. The $j$-th scalar integral in the $i$-th family is defined as
\begin{align}
\label{eq:def_integral_1}
F^{(1)}_i\left[\vec{a}_j\right] = \int \frac{d^dl_1}{(2\pi)^d} \, f^{(1)}_i\left[\vec{a}_j\right]\,, \qquad f^{(1)}_i\left[\vec{a}_j\right] = \frac{1}{D_{i,1}^{a_{j,1}} D_{i,2}^{a_{j,2}} D_{i,3}^{a_{j,3}} D_{i,4}^{a_{j,4}} D_{i,5}^{a_{j,5}}}\,, 
\end{align}
where $f^{(1)}_i\left[\vec{a}_j\right]$ denotes the integrand of the $j$-th one-loop Feynman integral in the $i$-th family, and $D_{i,k}$ ($k=1,\ldots,5$) are the corresponding propagator denominators. The choice of $D_{i,k}$ for each family is not unique. For instance, for the first family we define
\begin{align}
&T^{(1)}_1=\left\{l_1^2,\  (l_1-p_1)^2-m_t^2,\ (l_1-p_1-p_2)^2,\ (l_1-p_1-p_2+p_3)^2, \ (l_1-p_5)^2\right\} \,.
\end{align}
The definitions of all $17$ families are provided in ancillary electronic files attached with this paper. According to Eq.~\eqref{eq:matirxE0}, the NLO matrix elements of the bare squared amplitudes can be written as
\begin{align}\label{eq:sqamp1}
\left(\bm{H}^{(1)\text{bare}}_{q,g}\right)_{IJ}&= \left(d^{(0)\text{bare}}_{J,q,g}\right)^{\dag}d^{(1)\text{bare}}_{I,q,g} \nonumber \\
&=\int \frac{d^dl_1}{(2\pi)^d}\sum_{ij}c^{(1)\text{bare}}_{ij,IJ,q,g}f^{(1)}_i\left[\vec{a}_j\right]\,.
\end{align}

After applying IBP reduction with the package \texttt{Kira}~\cite{Maierhoefer:2017hyi,Klappert:2020nbg}, we identify $107$ master integrals (MIs), including mappings among the MIs of all families.\footnote{The MIs required for calculating the $n_f^2$ terms of the two-loop amplitude form a subset of these $107$ MIs.} Up to linear transformations arising from different definitions of integral families and choices of MIs, all of these integrals have already been evaluated numerically in the literature using the method of differential equations~\cite{Badger:2022mrb,Bera:2025upg}. In this work, we calculate all $107$ MIs numerically with the package \texttt{AMFlow}~\cite{Liu:2022chg}, which allows us to obtain the amplitudes up to $\mathcal{O}(\epsilon^2)$, as required for the NNLO calculation. We have also compared our NLO squared amplitudes up to $\mathcal{O}(\epsilon^0)$ with those produced by the package \texttt{GoSam}~\cite{GoSam:2014iqq}, and we find complete agreement, modulo the choice of dimensional regularization scheme.\footnote{We work in conventional dimensional regularization (CDR), where all quantities live in $d$ dimensions, while \texttt{GoSam} implements the 't Hooft-Veltman (HV) scheme, in which external legs remain in $4$ dimensions.}

\subsection{Massive amplitudes at two loops}\label{sec:massive_amp_2}

To obtain approximate expressions for the finite terms of the two-loop amplitudes using the mass-factorization approach~\cite{Mitov:2006xs} in the high-energy boosted limit, we require the amplitudes up to two-loop order with all parton masses set to zero. For generality, we first generate the two-loop amplitudes with full mass dependence using \texttt{FeynArts}~\cite{Hahn:2000kx} and process the expressions with \texttt{FeynCalc}~\cite{Shtabovenko:2020gxv,Shtabovenko:2023idz}. The amplitudes are then expressed in terms of scalar integrals by multiplying the two-loop amplitudes by the tree-level ones.

Similar to the one-loop case, the two-loop scalar integrals can be classified into $467$ families (before accounting for permutations of the external momenta). Each family is defined by $11$ propagator denominators. The $j$-th scalar integral in the $i$-th family is defined as
\begin{align}
\label{eq:def_integral_2}
F^{(2)}_i\left[\vec{a}_j\right] = \int \frac{d^dl_1}{(2\pi)^d}\frac{d^dl_2}{(2\pi)^d} f^{(2)}_i\left[\vec{a}_j\right] \,, \qquad f^{(2)}_i\left[\vec{a}_j\right]=\prod_{r=1}^{11}\frac{1}{D_{i,r}^{a_{j,r}} }\, , 
\end{align}
where $f^{(2)}_i\left[\vec{a}_j\right]$ denotes the integrand of the $j$-th two-loop scalar integral in the $i$-th family. For the first family, the propagator denominators are chosen as
\begin{align}
&T^{(2)}_1=\big\{l_1^2,\ \left(l_1-p_1\right)^2,\ \left(l_1+p_2-p_3-p_4\right)^2,\ \left(l_1-p_3-p_4\right)^2,\ l_2^2,\ \left(l_2-p_4\right)^2-m_t^2, \nonumber \\  
&\hspace{-6mm}\qquad\quad\ \left(l_1-l_2\right)^2,\  \left(l_1-p_4\right)^2-m_t^2,\ \left(l_2-p_1\right)^2,\ \left(l_2+p_2-p_3-p_4\right)^2,\ \left(l_2-p_3-p_4\right)^2 \big\} \,.
\end{align}
The definitions of all $467$ families are provided in  ancillary electronic files accompanying this paper. As a result, the matrix elements of the interference between $\mathcal{M}^{(2)}$ and $\mathcal{M}^{(0)}$ can be written as
\begin{align}\label{eq:sqamp2}
\left(\bm{H}^{(2)\text{bare}}_{q,g}\right)_{IJ}&= \left(d^{(0)\text{bare}}_{J,q,g}\right)^{\dag}d^{(2)\text{bare}}_{I,q,g} \nonumber \\
&= \int\frac{d^dl_1}{(2\pi)^d}\frac{d^dl_2}{(2\pi)^d}\sum_{ij}c^{(2)\text{bare}}_{ij,IJ,q,g}f^{(2)}_i\left[\vec{a}_j\right] \,.
\end{align}
Without further simplification, we find that roughly $1.7 \times 10^6$ two-loop scalar integrals $F^{(2)}_i\left[\vec{a}_j\right]$ are needed to express $\bm{H}^{(2)\text{bare}}_{q,g}$ for both the $q\bar{q}$ and $gg$ channels. These two-loop Feynman integrals involve six scales and are extremely challenging, rendering an exact calculation of the two-loop amplitudes for $t\bar{t}\gamma$ production beyond the reach of current methods. Apart from the IR singularities of the two-loop amplitudes, which can be predicted from Eq.~\eqref{eq:IRamps}, in what follows we employ the mass-factorization approach to approximate their finite terms.

\subsection{The massive amplitudes from mass factorization}
\label{sec:mass-fac}

In this section, we present approximate amplitudes for $t\bar{t}\gamma$ production in the high-energy boosted limit. To this end, we adopt the mass-factorization formula~\cite{Mitov:2006xs, Becher:2007cu, Engel:2018fsb, Wang:2023qbf}:
\begin{multline}
\label{eq:QCDfac1}
\Ket{\bar{\mC{M}}^R_{q,g}\lt(\epsilon,\{p\},m_t,\mu\rt)} = \zcal_{[q,g]}^{(m|0)}(\epsilon,m_t,\mu) \, \zcal_{[t]}^{(m|0)}(\epsilon,m_t,\mu)
\\
\times \bm{\scal}(\epsilon,\{\tilde{p}\},m_t,\mu)\Ket{\td{\mC{M}}^R_{q,g}\lt(\epsilon,\{\tilde{p}\},\mu\rt)} \,,
\end{multline}
where $\tilde{\mathcal{M}}^R$ is the renormalized massless amplitude with all parton masses set to zero, and $\bar{\mC{M}}^R$ approximates the full amplitude $\mC{M}^R$ up to power-suppressed corrections in $m_t^2$.\footnote{This factorization formula is valid at LP; sub-leading power contributions are under active investigation~\cite{Gervais:2017yxv, Laenen:2020nrt, Fadin:2023phc, vanBijleveld:2025ekz}.} The $\zcal$-factors $\zcal_{[j]}^{(m|0)}$ and the soft function $\bm{\scal}$ contain all logarithmic mass dependence, $\ln(\mu^2/m_t^2)$. General results for $\zcal_{[j]}^{(m|0)}$ and $\bm{\scal}$ up to NNLO can be found in Refs.~\cite{Mitov:2006xs, Czakon:2007ej, Czakon:2007wk, Becher:2007cu, Engel:2018fsb, Wang:2023qbf}, while the specific expressions relevant for $t\bar t \gamma$ coincide with those for $t\bar t H$ production and are given in Appendix~C of Ref.~\cite{Wang:2024pmv}. We use $\{p\}$  and $\{\tilde{p}\}$ to denote the sets of external momenta in the massive and massless cases, respectively. The momenta in $\tilde{\mathcal{M}}^R$ satisfy $\tilde{p}_i^2 = 0$, and we define $\tilde{s}_{ij} \equiv 2\sigma_{ij} \, \tilde{p}_i \cdot \tilde{p}_j$. In the high-energy boosted limit, $s_{ij}\approx\tilde{s}_{ij}$. There remains some freedom in how the massless momenta $\{\tilde{p}\}$ are related to the massive ones $\{p\}$. Before specifying our choice, we turn to the calculation of the two-loop massless amplitude $\tilde{\mathcal{M}}^R$.

Analogous to the massive case, we renormalize the massless amplitudes as 
\begin{align}\label{eq:renormalmassless}
\Ket{{\td{\mC{M}}}^R_{q,g}(\alpha_s, e, \{\tilde{p}\}, \mu, \epsilon)} = \left(\frac{\mu^2 e^{\gamma_E}}{4\pi}\right)^{-\epsilon} \Ket{\td{\mC{M}}^{\text{bare}}_{q,g}(\alpha_s^0, e^0, \{\tilde{p}\}, \epsilon)} \,.
\end{align}
In the massless case, the on-shell wave-function renormalization constants for gluons and quarks are unity, and the renormalization of $\alpha_s$ is the same as in Eq.~\eqref{eq:renorcoupling}. The bare massless amplitudes $\td{\mC{M}}^{\text{bare}}_{q,g}(\{\tilde{p}\})$ can be obtained by taking the massless limit of the massive ones $\mC{M}^{\text{bare}}_{q,g}(\{p\}, m_t^0)$ at the integrand level, i.e. before loop integration. Consequently, the N$^k$LO ($k=1,2$) matrix elements of the bare massless squared amplitudes can be written analogously to Eq.~\eqref{eq:sqamp2}:\footnote{Here we only consider the interference between the $k$-loop amplitude and the tree-level one.}
\begin{align}\label{eq:sqamp3}
\left(\tilde{\bm{H}}^{(k)\text{bare}}_{q,g}\right)_{IJ}= \int\left(\prod_{r=1}^{k}\frac{d^dl_r}{(2\pi)^d}\right)\sum_{ij}\tilde{c}^{(k)\text{bare}}_{ij,IJ,q,g}\tilde{f}^{(k)}_i\left[\vec{a}_j\right]\,,
\end{align}
with
\begin{align}\label{eq:transmass}
\tilde{c}^{(k)\text{bare}}_{ij,IJ,q,g}=\left.c^{(k)\text{bare}}_{ij,IJ,q,g}\right|_{m_t\to 0, \,p_i\to \tilde{p}_i} \,, \quad
\tilde{f}^{(k)}_i\left[\vec{a}_j\right]=\left.f^{(k)}_i\left[\vec{a}_j\right]\right|_{m_t\to 0, \,p_i\to \tilde{p}_i}\,.
\end{align}
Note that $\tilde{c}^{(k)\text{bare}}_{ij,IJ,q,g}$ and $\tilde{f}^{(k)}_i\left[\vec{a}_j\right]$ are exactly the LP contributions of $c^{(k)\text{bare}}_{ij,IJ,q,g}$ and $f^{(k)}_i\left[\vec{a}_j\right]$ in the high-energy boosted limit, respectively.

The NNLO massless squared amplitudes can therefore be expressed as linear combinations of two-loop massless scalar Feynman integrals. In this case, the number of two-loop scalar integrals is reduced to about $7.8 \times 10^5$, compared to the massive case. These scalar integrals are categorized into $187$ integral families, which reduce to four independent topologies after accounting for permutations of external momenta. We carry out the IBP reduction using the package \texttt{Kira}~\cite{Maierhoefer:2017hyi, Klappert:2020nbg} with the help of \texttt{FireFly}~\cite{Klappert:2019emp, Klappert:2020aqs}, expressing all scalar integrals in terms of master integrals\footnote{Most reduction relations have been obtained in Ref.~\cite{Wang:2024pmv}; further details can be found there.} studied in Refs.~\cite{Gehrmann:2015bfy,Papadopoulos:2015jft,Gehrmann:2018yef,Chicherin:2018mue,Abreu:2018rcw,Abreu:2018aqd,Chicherin:2018old,Badger:2019djh}, which can be evaluated numerically using \texttt{PentagonMI}~\cite{Chicherin:2020oor}.\footnote{The MIs obtained by IBP reduction must be further expressed in terms of the so-called uniform transcendentality (UT) basis chosen in Ref.~\cite{Chicherin:2020oor} via linear transformations.} Using the same program, we also compute the one-loop UT bases up to weight $4$ required for the one-loop amplitudes up to $\mathcal{O}\left(\epsilon^2\right)$. 
Finally, we validate the massless squared amplitudes by checking their IR poles against those predicted by the massless IR subtraction factors~\cite{Becher:2009cu} and the corresponding lower-order squared amplitudes. We find excellent agreement within numerical accuracy.

\section{Approximate massive squared amplitudes and mapping between massive and massless phase-space points}\label{sec:squaredamp}

As noted in Eq.~\eqref{eq:QCDfac1}, the factorization formula is valid at LP and receives power-suppressed corrections. Consequently, ambiguities of $\mathcal{O}(m_t^2)$ arise, depending on the specific convention adopted to implement the factorization. In particular, when applying the formula to compute physical cross sections, two types of ambiguities occur, which we discuss in this section.

\subsection{Approximate massive squared amplitudes}

The first ambiguity concerns how the factorization formula is applied to compute squared amplitudes. Since the tree-level and one-loop massive amplitudes can be calculated exactly, the only ambiguity in the NNLO squared amplitudes lies in the interference between the two-loop and tree-level amplitudes. To validate our approximate approach, we also present results at NLO in the high-energy boosted limit. We consider two representative schemes and compare their outcomes, while keeping in mind that other choices are possible.

In the first scheme, which we refer to as the ``massless scheme'', we directly square the approximate amplitudes $\ket{\bar{\mC{M}}^R_{q,g}}$ given in Eq.~\eqref{eq:QCDfac1} and extract the NLO and NNLO terms. Here we only consider the interferences of the amplitudes with the tree-level ones, namely $\braket{\bar{\mC{M}}^{(0)R}_{q,g}|\bar{\mC{M}}^{R}_{q,g}}$. Note that $\ket{\bar{\mC{M}}^{(0)R}_{q,g}}$ are identical to the massless tree-level amplitudes $\ket{\td{\mC{M}}^{(0)R}_{q,g}}$. All mass dependence is contained in the factors $\zcal_{[j]}^{(m|0)}$ and the soft function $\bm{\scal}$. Combining Eqs.~\eqref{eq:sqamp0},~\eqref{eq:QCDfac1},~\eqref{eq:renormalmassless}, and~\eqref{eq:sqamp3}, we obtain
\begin{align}\label{eq:massless_scheme}
\braket{\bar{\mC{M}}^{(0)R}_{q,g}|\bar{\mC{M}}^{R}_{q,g}}&=\zcal_{[q,g]}^{(m|0)}(\epsilon,m_t,\mu) \, \zcal_{[t]}^{(m|0)}(\epsilon,m_t,\mu) \nonumber \\
&\times \sum_{IJK}\Big[\bm{\scal}(\epsilon,\{\tilde{p}\},m_t,\mu)\Big]^{q,g}_{KI}\Big[\tilde{\bm{H}}^R_{q,g}(\epsilon,\{\tilde{p}\},\mu)\Big]_{IJ}\braket{c_J|c_K} \,,
\end{align}
where
\begin{align}\label{eq:softfunctionME}
\bm{\scal}^{q,g}_{KI} = \frac{\Bra{c^{q,g}_K}\bm{\scal}\Ket{c^{q,g}_I}}{\braket{c^{q,g}_K|c^{q,g}_K}}\,.
\end{align}
Although the soft function $\bm{\scal}$ has the same operator form in both the $q\bar{q}$ and $gg$ channels, the matrix elements in Eq.~\eqref{eq:softfunctionME} differ between the partonic channels due to their distinct color structures~\cite{Wang:2023qbf}.
To extract the NLO and NNLO contributions, we require $\tilde{\bm{H}}^R_{q,g}$ up to NNLO. At LO,
\begin{align}
\Big[\tilde{\bm{H}}^{(0)R}_{q,g}\Big]_{IJ} = \Big[\tilde{\bm{H}}^{(0)\text{bare}}_{q,g}\Big]_{IJ} =\left(\tilde{d}^{(0)\text{bare}}_{J,q,g}\right)^{\dag}\tilde{d}^{(0)\text{bare}}_{I,q,g}\,,
\end{align}
and the higher order terms $\tilde{\bm{H}}^{(k)R}_{q,g}$ ($k=1,2$) follow from Eq.~\eqref{eq:sqamp3} after UV renormalization, which are given by
\begin{align}
\tilde{\bm{H}}^{(1)R}_{q,g} &= \left(\frac{\mu^2e^{\gamma_E}}{4\pi}\right)^{\epsilon}\tilde{\bm{H}}^{(1)\text{bare}}_{q,g} + Z^{(1)}_{\alpha_s}\tilde{\bm{H}}^{(0)\text{bare}}_{q,g}\,, \nonumber \\
\tilde{\bm{H}}^{(2)R}_{q,g} &= \left(\frac{\mu^2e^{\gamma_E}}{4\pi}\right)^{2\epsilon}\tilde{\bm{H}}^{(2)\text{bare}}_{q,g} + 2\left(\frac{\mu^2e^{\gamma_E}}{4\pi}\right)^{\epsilon}Z^{(1)}_{\alpha_s}\tilde{\bm{H}}^{(1)\text{bare}}_{q,g} + Z^{(2)}_{\alpha_s}\tilde{\bm{H}}^{(0)\text{bare}}_{q,g}\,.
\end{align}

The second scheme is slightly more involved. We introduce a modified version of Eq.~\eqref{eq:massless_scheme}:
\begin{align}\label{eq:massive_scheme}
\braket{\hat{\mC{M}}^{(0)R}_{q,g}|\hat{\mC{M}}^{R}_{q,g}}&=\zcal_{[q,g]}^{(m|0)}(\epsilon,m_t,\mu) \, \zcal_{[t]}^{(m|0)}(\epsilon,m_t,\mu) \nonumber \\
&\times \sum_{IJK}\Big[\bm{\scal}(\epsilon,\{\tilde{p}\},m_t,\mu)\Big]_{KI}\Big[\hat{\bm{H}}^R_{q,g}(\epsilon,\{\tilde{p}\},\{p\},m_t,\mu)\Big]_{IJ}\braket{c_J|c_K}\,,
\end{align}
where the modified matrices $\hat{\bm{H}}^R_{q,g}$ are defined up to NNLO as
\begin{align}
\Big[\hat{\bm{H}}^{(0)R}_{q,g}\Big]_{IJ} &= \Big[\bm{H}^{(0)R}_{q,g}\Big]_{IJ} =\left(d^{(0)\text{bare}}_{J,q,g}\right)^{\dag}d^{(0)\text{bare}}_{I,q,g}\,,\nonumber \\
\hat{\bm{H}}^{(1)R}_{q,g} &= \left.\left(\frac{\mu^2e^{\gamma_E}}{4\pi}\right)^{\epsilon}\bm{H}^{(1)\text{bare}}_{q,g}\right|_{f^{(1)}_i\left[\vec{a}_j\right]\to \tilde{f}^{(1)}_i\left[\vec{a}_j\right]} + Z^{(1)}_{\alpha_s}\bm{H}^{(0)\text{bare}}_{q,g}\,, \nonumber \\
\hat{\bm{H}}^{(2)R}_{q,g} &= \left.\left(\frac{\mu^2e^{\gamma_E}}{4\pi}\right)^{2\epsilon}\bm{H}^{(2)\text{bare}}_{q,g}\right|_{f^{(2)}_i\left[\vec{a}_j\right]\to \tilde{f}^{(2)}_i\left[\vec{a}_j\right]} + \left.2\left(\frac{\mu^2e^{\gamma_E}}{4\pi}\right)^{\epsilon}Z^{(1)}_{\alpha_s}\bm{H}^{(1)\text{bare}}_{q,g}\right|_{f^{(1)}_i\left[\vec{a}_j\right]\to \tilde{f}^{(1)}_i\left[\vec{a}_j\right]} \nonumber \\
&+ Z^{(2)}_{\alpha_s}\bm{H}^{(0)\text{bare}}_{q,g}\,,
\end{align}
with $\bm{H}^{(1)\text{bare}}_{q,g}$ and $\bm{H}^{(2)\text{bare}}_{q,g}$ defined in Eqs.~\eqref{eq:sqamp1} and~\eqref{eq:sqamp2}, respectively. The massive integrands $f^{(1)}_i\left[\vec{a}_j\right]$ and $f^{(2)}_i\left[\vec{a}_j\right]$ are given in Eqs.~\eqref{eq:def_integral_1} and~\eqref{eq:def_integral_2}, while their massless counterparts $\tilde{f}^{(k)}_i\left[\vec{a}_j\right]$ are defined in Eq.~\eqref{eq:transmass}. In other words, the modified matrices $\hat{\bm{H}}^R_{q,g}$ are obtained by replacing all scalar integrals in the bare massive matrices $\bm{H}^{\text{bare}}_{q,g}$ with their massless counterparts, and then performing UV renormalization as in the massless case. Thus, the only difference between Eqs.~\eqref{eq:massive_scheme} and~\eqref{eq:massless_scheme} arises from the distinction between $\hat{\bm{H}}^R_{q,g}$ and $\tilde{\bm{H}}^R_{q,g}$, which is power suppressed. In this setup, $\hat{\bm{H}}^R_{q,g}$ retains full mass information from the external top-quark pair and partial mass information from internal top quark propagators. We therefore refer to this scheme as the ``semi-massive scheme''.

\subsection{Mapping between massive and massless phase-space points}

The second ambiguity concerns the relation between the massive external momenta $p_i$ and the corresponding massless ones $\tilde{p}_i$ in Eq.~\eqref{eq:QCDfac1}. Following the prescription used for $t\bar{t}H$ production~\cite{Wang:2024pmv}, we keep the directions of the 3-momenta identical between $p_i$ and $\tilde{p}_i$ and rescale their magnitudes (and energy components) such that the massless on-shell conditions and momentum conservation are both satisfied.

Specifically, since $p_1$ and $p_2$ are already massless, we simply set $\tilde{p}_1=p_1$ and $\tilde{p}_2=p_2$ resulting in $\td{s}_{12}=s_{12}$. We parameterize the massive external momenta in the center-of-mass frame as
\begin{align}\label{eq:momentamassive}
p_1&=\frac{\sqrt{s_{12}}}{2}\lt(1,0,0,1\rt)\,,\nn\\
p_2&=\frac{\sqrt{s_{12}}}{2}\lt(1,0,0,-1\rt)\,,\nn\\
p_3&=\lt(\sqrt{m_t^2+q_3^2},q_3\sin\theta_3\sin\phi_3,q_3\sin\theta_3\cos\phi_3,q_3\cos\theta_3\rt)\,,\nn\\
p_4&=\lt(\sqrt{m_t^2+q_4^2},q_4\sin\theta_4\sin\phi_4,q_4\sin\theta_4\cos\phi_4,q_4\cos\theta_4\rt)\,,\nn\\
p_5&=q_5\lt(1,\sin\theta_5,0,\cos\theta_5\rt) \,,
\end{align}
where $q_i$ is the magnitude of the spatial component of $p_i$, $\theta_i$ the polar angle, and $\phi_i$ the azimuthal angle. For convenience and without loss of generality, we have set $\phi_5=0$.
Momentum conservation and on-shell conditions leave six independent parameters on the right-hand side of Eq.~\eqref{eq:momentamassive}, which we choose as $s_{12}$, $m_t$, $q_5$, $\theta_3$, $\phi_3$, and $\theta_5$. Their values satisfy the physical constraints
\begin{align}
&s_{12} \ge 4m_t^2\,, \quad 0\le q_5\le q_{5,\rm max}=\frac{s_{12}-4m_t^2}{2\sqrt{s_{12}}}\,,\nn \\
m_t >&0 \,, \quad 0\le \theta_3 \le \pi\,, \quad 0\le \phi_3 < 2\pi\,, \quad 0\le \theta_5 \le \pi \,.
\end{align}

The massless momenta $\tilde{p}_i$ are then obtained by rescaling $\tilde{p}_{3,4,5}$ at fixed angles to enforce $\tilde{p}_i^2=0$ and momentum conservation, following the same prescription as in $t\bar{t}H$ production~\cite{Wang:2024pmv}.

\section{Numeric results}\label{sec:numeric}

We are now ready to present numerical results for the two-loop amplitudes. For the input parameters, we take the top-quark mass $m_t = \SI{173}{\GeV}$ and, by default, set the renormalization scale to $\mu = m_t$. Other choices of $\mu$ are of course possible and will be considered below.\footnote{In the high-energy boosted region, where the kinematic scales satisfy $|s_{ij}| \gg m_t^2$, large logarithms such as $\ln(-s_{ij}/m_t^2)$ appear regardless of the choice of $\mu$. The standard way to address this problem is to resum these logarithms using renormalization group equations, based on the factorization formula \eqref{eq:QCDfac1} (see, e.g., \cite{Mele:1990cw, Ferroglia:2012ku, Ferroglia:2013awa, Pecjak:2016nee, Czakon:2018nun}). In this work, however, we use the factorization formula only to approximate the two-loop amplitudes and are therefore free to choose any scale $\mu$ when presenting our results.}

In addition, the amplitudes depend on the electric charges of the quarks, entering both through the external states and through closed internal quark loops. As a result, in the $q\bar{q}$ channel, we must distinguish between up-type (e.g., $u\bar{u}$) and down-type (e.g., $d\bar{d}$) initial states. For contributions from closed light-quark loops with a photon attached, we likewise need to distinguish between up-type quarks with electric charge $2/3$ and down-type quarks with charge $-1/3$. In both cases the contributions from down-type light quarks can be obtained from those of up-type light quarks by including a factor of $-1/2$ for each $q\bar{q}\gamma$ vertex, without introducing additional computational complexity. For convenience, and without loss of generality, we therefore take the electric charge of all quarks to be $Q_f = 2/3$ when presenting numerical results, which is expected to have only a mild impact on assessing the validity of our approximate approach.

To test the validity of the approximation, we first compare the approximate results with the exact ones. The exact results include the NLO squared amplitudes up to $\mathcal{O}(\epsilon^2)$, and the IR poles of the NNLO squared amplitudes predicted by Eq.~\eqref{eq:IRamps}. The NNLO squared amplitudes receive two types of contributions: the one-loop-squared amplitudes, which can be obtained exactly, and the interference between the two-loop and tree-level amplitudes. For clarity, we therefore drop the one-loop-squared contribution in the comparison and focus only on the interference terms.

\begin{figure}[t!]
\centering  
\subfigure{
\label{fig:qqnloeps0s}
\includegraphics[width=0.48\textwidth]{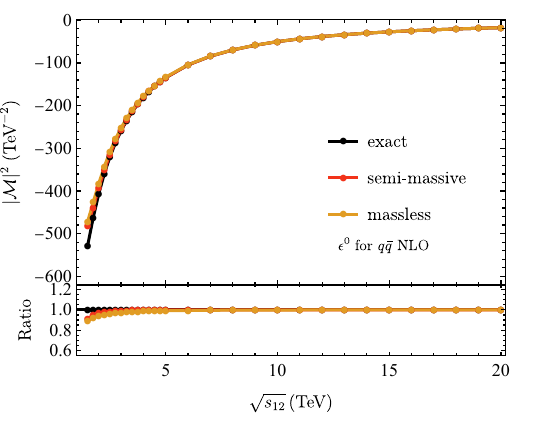}}
\subfigure{
\label{fig:qqnloepsm1s}
\includegraphics[width=0.48\textwidth]{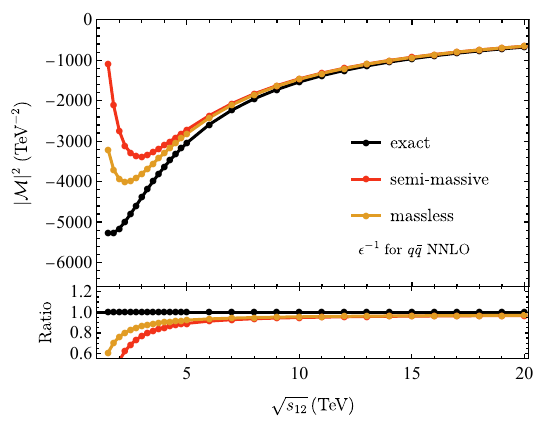}} 
\subfigure{
\label{fig:ggnnloeps0s}
\includegraphics[width=0.48\textwidth]{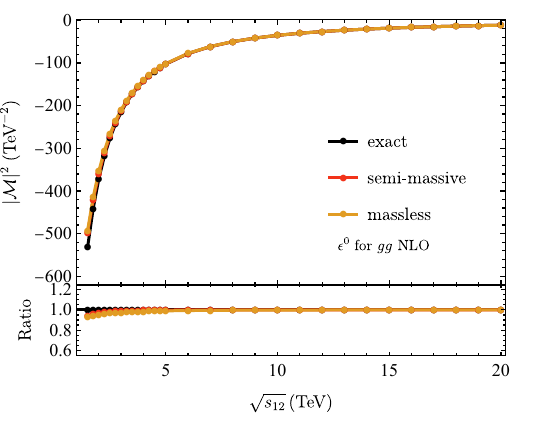}}
\subfigure{
\label{fig:ggnnloepsm1s}
\includegraphics[width=0.48\textwidth]{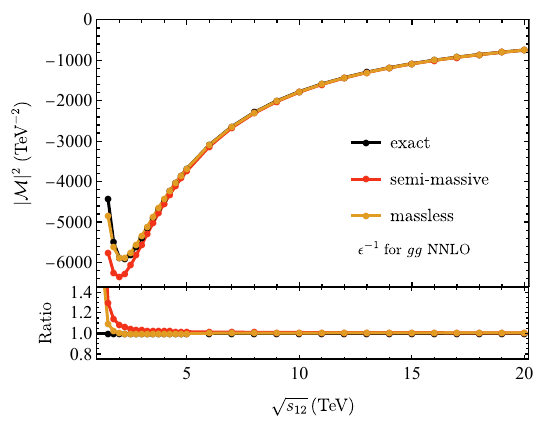}}
\caption{The squared amplitudes at $\mC{O}(\epsilon^0)$ for NLO and $\mC{O}(\epsilon^{-1})$ for NNLO in both the $q\bar{q}$ and $gg$ channels are shown as functions of the partonic center-of-mass energy $\sqrt{s_{12}}$. The other kinematic variables are fixed to $\theta_3 = 14\pi/29$, $\phi_3=34\pi/29$, $\theta_5=15\pi/29$ and $q_5=20\,q_{5,\rm max}/29$. Spin- and color-average factors of $1/36$ ($q\bar{q}$) and $1/256$ ($gg$) are included. The lower panel in each plot shows the ratios of the approximate results to the exact ones.}
\label{fig:diffscheme}
\end{figure}

We first show in Fig.~\ref{fig:diffscheme} the numerical results for the squared amplitudes at $\mC{O}(\eps^0)$ (NLO, left plots) and $\mC{O}(\eps^{-1})$ (NNLO, right plots) in both the $q\bar{q}$ (upper plots) and $gg$ (lower plots) channels. The partonic center-of-mass energy $\sqrt{s_{12}}$ is varied from $\SI{1.5}{\TeV}$ to $\SI{20}{\TeV}$, while the remaining independent kinematic variables are fixed as $\theta_3 = 14\pi/29$, $\phi_3=34\pi/29$, $\theta_5=15\pi/29$, and $q_5=20\,q_{5,\rm max}/29$. The black lines represent the exact results, while the red and orange curves denote the approximate results obtained using the two different schemes discussed in Sec.~\ref{sec:squaredamp}. As expected, the approximation improves with increasing energy for both NLO and NNLO. We observe that the semi-massive scheme (red curves) performs slightly better than the massless scheme (orange curves) at NLO, whereas at NNLO the opposite is true. This behavior can be traced to the fact that the semi-massive scheme incorporates part of the power-suppressed contributions ($\mathcal{O}(m_t^2)$ and higher). However, this should be treated with caution, since beyond-LP terms can only be reliable once the unknown power-suppressed contributions from the scalar Feynman integrals are included. Overall, the massless scheme provides a reasonable approximation to the exact results down to $\sqrt{s_{12}} \sim \SI{2}{\TeV}$ in all cases.

While the high-energy behavior is consistent with expectations, the different performance in the relatively low-energy region is more intriguing. From Fig.~\ref{fig:diffscheme}, we observe that at NLO the approximation behaves similarly in both the $gg$ and $q\bar{q}$ channels, whereas at NNLO it performs significantly better in the $gg$ channel. In the $q\bar{q}$ case, the relative deviation between the massless-scheme approximation and the exact result is much larger than the naive estimate of order $m_t^2/|s_{ij}|$. Moreover, the difference between the semi-massive and massless schemes is also non-negligible. These effects can be attributed to accidental cancellations at LP in the small-mass limit, which making beyond-LP contributions more important than anticipated. A similar effect already appears at LO: the LO squared amplitude is a rational function of $m_t^2$ and $s_{ij}$ and can be straightforwardly expanded in the small-mass limit. We find cancellations at LP for the contributions from a subset of Feynman diagrams. Although these diagrams are not dominant at LO, their impact can be enhanced at NNLO due to the complicated two-loop effects. In contrast, in the $gg$ channel, the relative difference between approximate and exact results is much smaller than what we expected, especially in the intermediate range $\SI{2}{\TeV} \lesssim \sqrt{s_{12}} \lesssim \SI{5}{\TeV}$. This improvement originates from another accidental cancellation, which will become evident in the color decomposition of the squared amplitude defined in Eq.~\eqref{eq:color_dec} in the following. As shown in Table~\ref{tab:ggnum1}, the ratio of the approximate result in the massless scheme to the exact result for the squared amplitude as a whole is significantly smaller than the corresponding ratios for each individual color configuration in the $\epsilon^{-1}$ column, owing to cancellations among different color coefficients.

\begin{figure}[t!]
\centering  
\subfigure{
\label{fig:qqnloeps0theta3}
\includegraphics[width=0.48\textwidth]{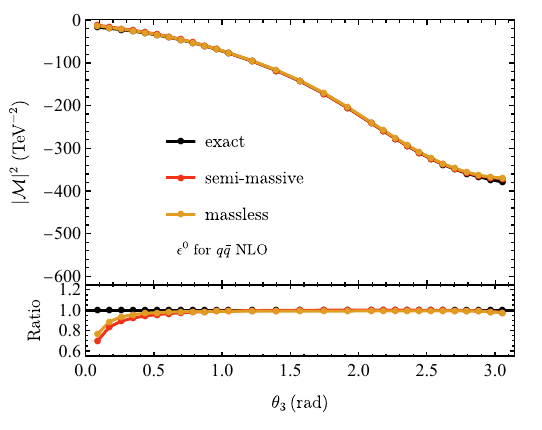}}
\subfigure{
\label{fig:qqnnloepsm1theta3}
\includegraphics[width=0.48\textwidth]{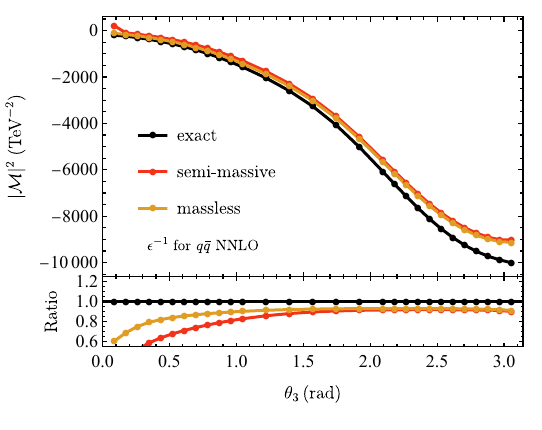}}
\subfigure{
\label{fig:ggnloeps0theta3}
\includegraphics[width=0.48\textwidth]{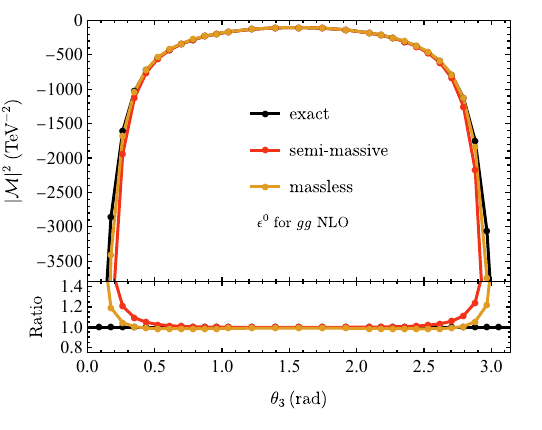}}
\subfigure{
\label{fig:ggnnloepsm1theta3}
\includegraphics[width=0.48\textwidth]{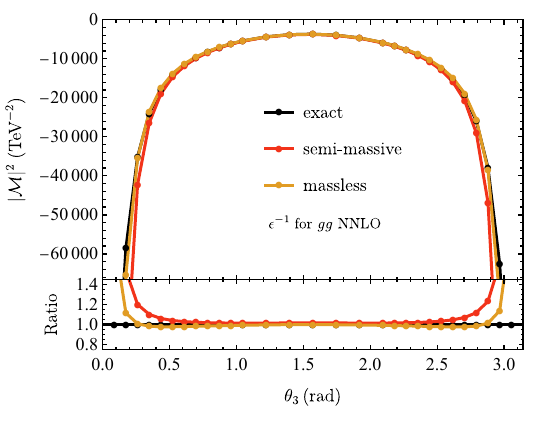}}
\caption{The squared amplitudes at $\mC{O}(\epsilon^0)$ for NLO and $\mC{O}(\epsilon^{-1})$ for NNLO in both the $q\bar{q}$ and $gg$ channels, shown as functions of the angle parameter $\theta_3$. The other kinematic variables are fixed as $\sqrt{s_{12}}=\SI{5}{\TeV}$, $\phi_3=34\pi/29$, $\theta_5=15\pi/29$ and $q_5=20\,q_{5,\rm max}/29$.}
\label{fig:diffschemetheta3s5}
\end{figure} 

We now turn to investigate the behavior of the approximate results as a function of the angle $\theta_3$ between $\vec{p}_3$ (the spatial momentum of the top quark) and $\vec{p}_1$. The other parameters are fixed to $\sqrt{s_{12}}= \SI{5}{\TeV}$, $\phi_3=34\pi/29$, $\theta_5=15\pi/29$ and $q_5=20\,q_{5,\rm max}/29$. As expected, when $\theta_3 \to 0$ or $\pi$, the approximation becomes less reliable due to the smallness of either $|s_{13}|$ or $|s_{23}|$. Indeed, Fig.~\ref{fig:diffschemetheta3s5} shows that the approximate results deviate significantly from the exact ones at the two ends of the spectrum. Especially in the $gg$ channel, the tree-level amplitudes contain propagators of the form $1/(s_{13}-m_t^2)$ and $1/(s_{23}-m_t^2)$, for which taking $m_t \to 0$ introduces sizable differences. This explains the wild behavior of the orange and red curves in the $gg$ channel near the two endpoints. Nevertheless, the massless scheme still provides a reasonable approximation across a large portion of the phase space in both the $q\bar{q}$ and $gg$ cases.

\begin{figure}[t!]
\centering  
\subfigure{
\label{fig:qqnloeps0thetah5}
\includegraphics[width=0.48\textwidth]{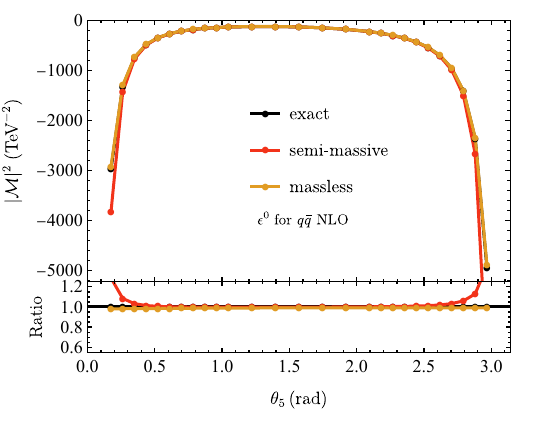}}
\subfigure{
\label{fig:qqnnloepsm1thetah5}
\includegraphics[width=0.48\textwidth]{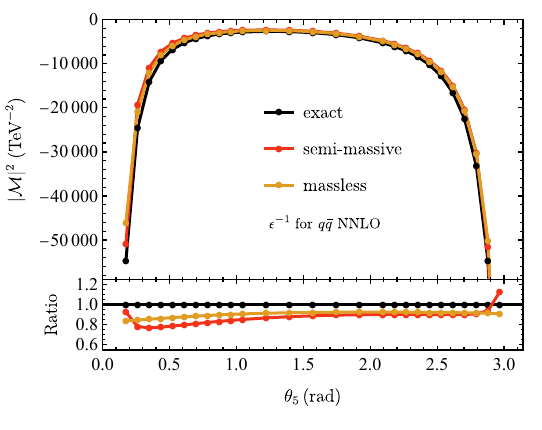}}
\subfigure{
\label{fig:ggnloeps0thetah5}
\includegraphics[width=0.48\textwidth]{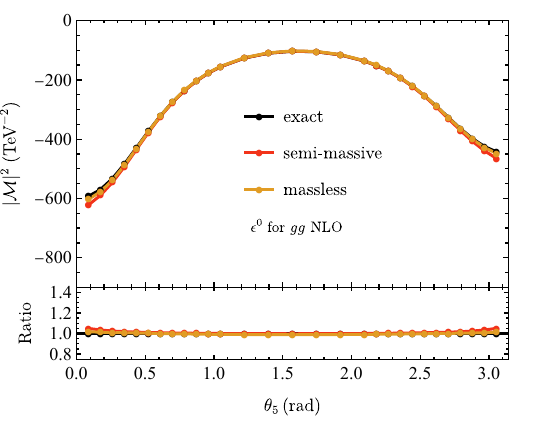}}
\subfigure{
\label{fig:ggnnloepsm1thetah5}
\includegraphics[width=0.48\textwidth]{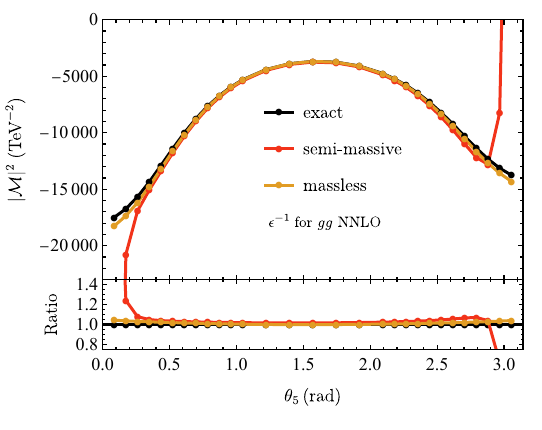}}
\caption{The squared amplitudes as a function of the angle parameter $\theta_5$. The other kinematic variables are chosen as: $\sqrt{s_{12}}=\SI{5}{\TeV}$, $\theta_3=14\pi/29$, $\phi_3=34\pi/29$ and $q_5=20\,q_{5,\rm max}/29$.}
\label{fig:diffschemethetah5s5}
\end{figure}

The situation is slightly different in the case of $\theta_5$, the angle between $\vec{p}_5$ (space configuration of the photon momentum) and $\vec{p}_1$. Since the photon is color neutral, QCD corrections are expected to have only a mild dependence on its momentum, which is indeed observed in Fig.~\ref{fig:diffschemethetah5s5}. Although collinear singularities arise at the two endpoints due to photon emission from the initial quark and antiquark (resulting in propagators $1/s_{15}$ and $1/s_{25}$) in the $q\bar{q}$ channel, the approximation in the massless scheme remains quite good across the whole range of $\theta_5$, provided that $\theta_3$ lies in the wide-angle region. By contrast, in the $gg$ channel at NNLO the red curve displays weird behavior, indicating large numerical cancellations between the power-suppressed terms included in the semi-massive scheme and the unknown contributions from scalar integrals. Taken together with the previous observations, this indicates that the massless scheme performs more reliably overall than the semi-massive scheme. We therefore adopt the massless scheme as the default for presenting numerical results in what follows.

\begin{figure}[t!]
\centering  
\subfigure{
\includegraphics[width=0.48\textwidth]{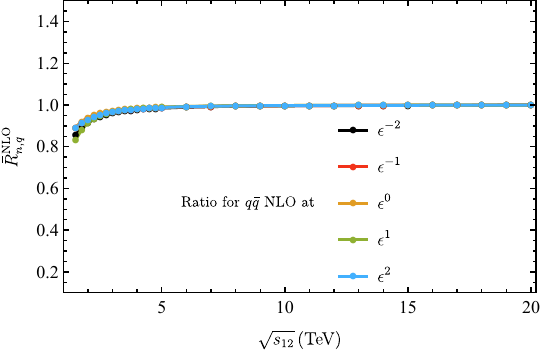}}
\subfigure{
\includegraphics[width=0.48\textwidth]{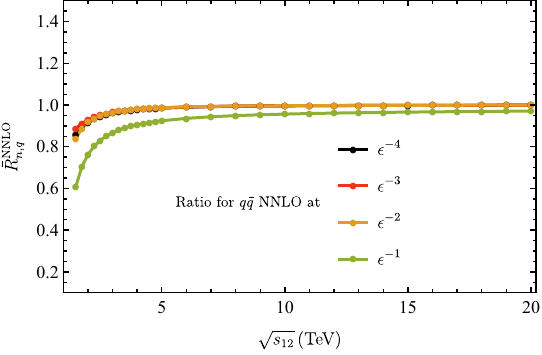}}
\subfigure{
\includegraphics[width=0.48\textwidth]{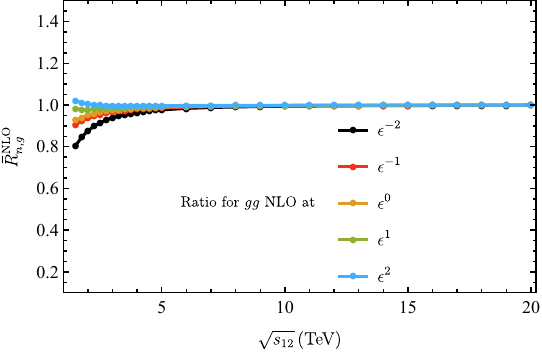}}
\subfigure{
\includegraphics[width=0.48\textwidth]{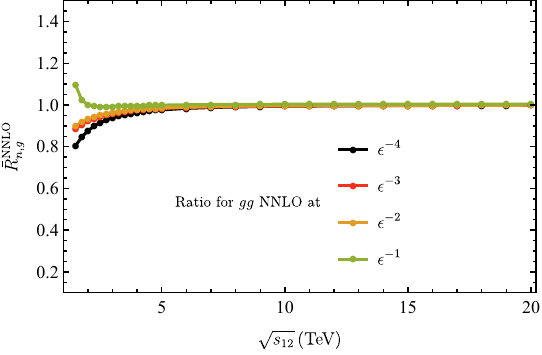}}
\caption{Ratios of the approximate results in the massless scheme to the exact ones at various orders in $\epsilon$, as functions of $\sqrt{s_{12}}$. The phase-space parameters are the same as those in Fig.~\ref{fig:diffscheme}.}
\label{fig:rationloshat}
\end{figure}

\begin{figure}[t!]
\centering  
\subfigure{
\includegraphics[width=0.48\textwidth]{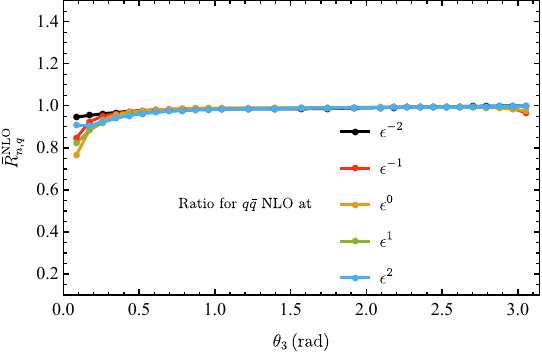}}
\subfigure{
\includegraphics[width=0.48\textwidth]{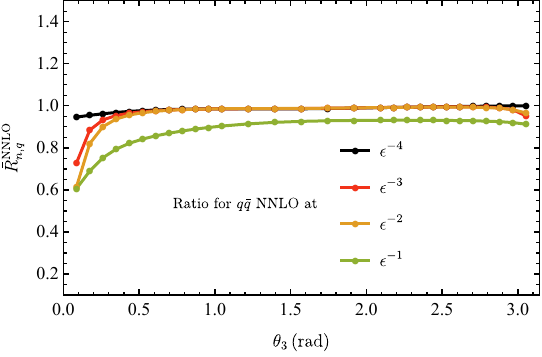}}
\subfigure{
\includegraphics[width=0.48\textwidth]{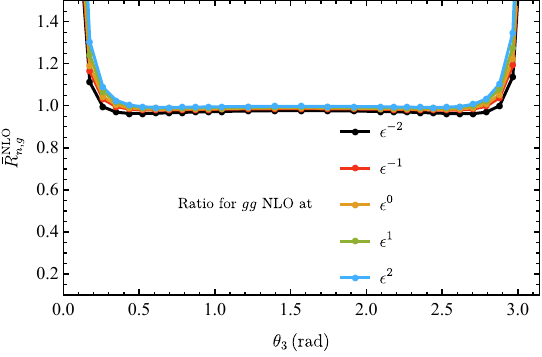}}
\subfigure{
\includegraphics[width=0.48\textwidth]{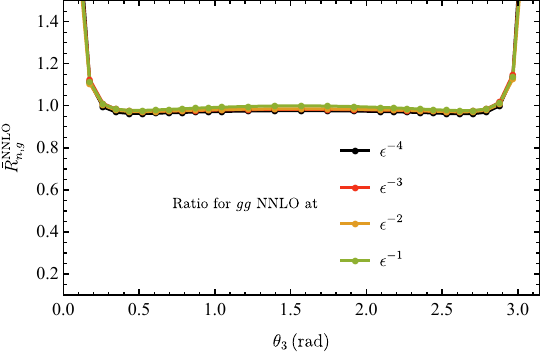}}
\caption{Ratios of the approximate results in the massless scheme to the exact ones at various orders in $\epsilon$, as functions of $\theta_3$. The other phase-space parameters are the same as those in Fig.~\ref{fig:diffschemetheta3s5}.}
\label{fig:rationlotheta5}
\end{figure}

We now consider the NLO and NNLO squared amplitudes at various orders in $\epsilon$ in the massless scheme. We define the following ratios of the approximate results to the exact ones at each order in $\epsilon$:
\begin{align}
\bar{R}_{n,q/g}^{\text{NLO}}=\frac{\mathrm{Re}\braket{\bar{\mC{M}}^{(0)R}_{q/g}|\bar{\mC{M}}^{(1)R}_{q/g}}\bigg|_{\epsilon^n}}{\mathrm{Re}\braket{\mC{M}^{(0)R}_{q/g}|\mC{M}^{(1)R}_{q/g}}\bigg|_{\epsilon^n}}\,, \quad \bar{R}_{n,q/g}^{\text{NNLO}}=\frac{\mathrm{Re}\braket{\bar{\mC{M}}^{(0)R}_{q/g}|\bar{\mC{M}}^{(2)R}_{q/g}}\bigg|_{\epsilon^n}}{\mathrm{Re}\braket{\mC{M}^{(0)R}_{q/g}|\mC{M}^{(2)R}_{q/g}}\bigg|_{\epsilon^n}}\,,
\end{align}
where $\ket{\bar{\mC{M}}^{(k)R}_{q/g}}$ are the amplitudes in the massless scheme, defined in Eq.~\eqref{eq:QCDfac1}, while $\ket{\mC{M}^{(k)R}_{q/g}}$ denote the exact massive amplitudes. $\mathrm{Re}\braket{\bar{\mC{M}}^{(0)R}_{q/g}|\bar{\mC{M}}^{(k)R}_{q/g}}$ are evaluated using Eq.~\eqref{eq:massless_scheme} in combination with Eq.~\eqref{eq:sqamp3}. The exact squared amplitudes $\mathrm{Re}\braket{\mC{M}^{(0)R}_{q/g}|\mC{M}^{(k)R}_{q/g}}$ are known up to $\epsilon^2$ at NLO and are computed up to $\epsilon^{-1}$ at NNLO in this work.

We show $\bar{R}_{n,q/g}^{\text{NLO}}$ and $\bar{R}_{n,q/g}^{\text{NNLO}}$ as functions of $\sqrt{s_{12}}$ in Fig.~\ref{fig:rationloshat}, using the same phase-space parameters as in Fig.~\ref{fig:diffscheme}. The approximation performs similarly across different coefficients in $\epsilon$ for both channels, except for the $\epsilon^{-1}$ term in the $q\bar{q}$ channel at NNLO (already noted in Fig.~\ref{fig:diffscheme}), and for the leading $1/\epsilon$ poles in the $gg$ channel at both NLO and NNLO. As implied by Eq.~\eqref{eq:IRamps}, the leading pole terms of the NLO and NNLO squared amplitudes are proportional to the LO ones as $\epsilon \to 0$. We indeed observe an accidental cancellation at LP in the small-mass limit for the LO squared amplitudes in the $gg$ channel at $\mathcal{O}(\epsilon^0)$, while no such cancellation occurs at higher orders in $\epsilon$.
In Fig.~\ref{fig:rationlotheta5}, we show $\bar{R}_{n,q,g}^{\text{NLO}}$ and $\bar{R}_{n,q,g}^{\text{NNLO}}$ as functions of $\theta_3$, using the same phase-space points as in Fig.~\ref{fig:diffschemetheta3s5}. The approximation remains reliable across most values of $\theta_3$ in the $q\bar{q}$ channel, except near $\theta_3 \to 0$, where the boosted limit is not valid due to small $|s_{13}|$. In the $gg$ channel, the approximation works very well in the central region but breaks down in the forward regions $\theta_3 \to 0$ or $\pi$, as expected.

\begin{figure}[t!]
\centering  
\subfigure{
\includegraphics[width=0.48\textwidth]{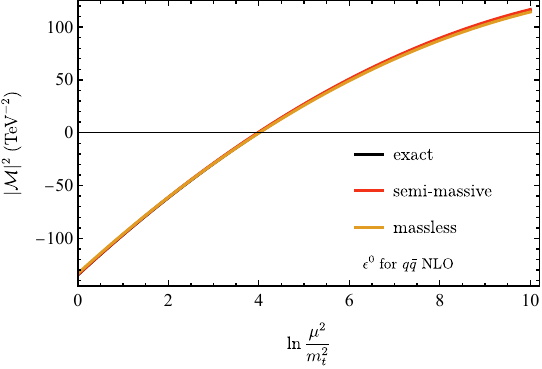}}
\subfigure{
\includegraphics[width=0.48\textwidth]{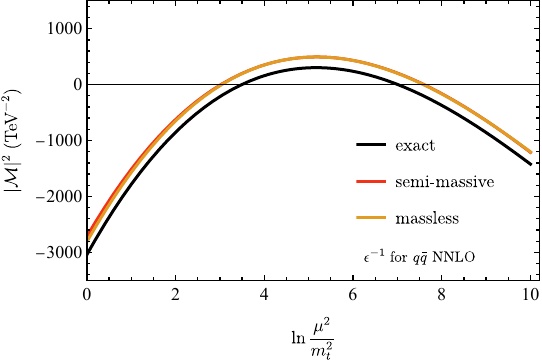}}
\subfigure{
\includegraphics[width=0.48\textwidth]{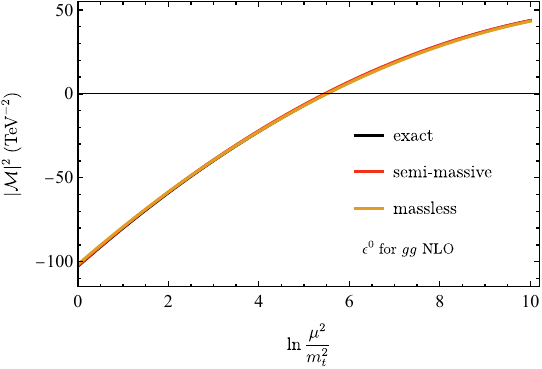}}
\subfigure{
\includegraphics[width=0.48\textwidth]{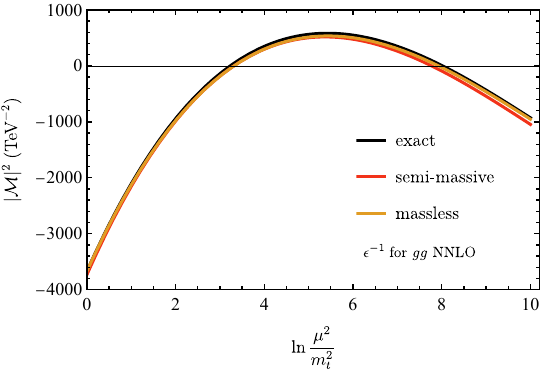}}
\caption{The squared amplitudes are shown as functions of the renormalization scale $\mu$. We show the $\epsilon^0$ coefficients at NLO and the $\epsilon^{-1}$ coefficients at NNLO. The center-of-mass energy is fixed to $\sqrt{s_{12}}=\SI{5}{\TeV}$ and the remaining phase-space parameters are the same as those in Fig.~\ref{fig:diffscheme}.}
\label{fig:squaredamplogmu5TeV}
\end{figure}

The above results were obtained using the default scale choice $\mu = m_t$. In Fig.~\ref{fig:squaredamplogmu5TeV}, we show the dependence of the squared amplitudes on the renormalization scale at NLO and NNLO. The approximation remains stable across different values of $\mu$, indicating that our formula correctly captures the scale dependence of the amplitudes.

\begin{figure}[t!]
\centering  
\subfigure{
\includegraphics[width=0.48\textwidth]{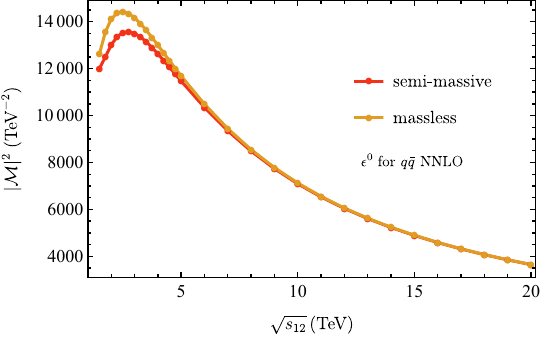}}
\subfigure{
\includegraphics[width=0.48\textwidth]{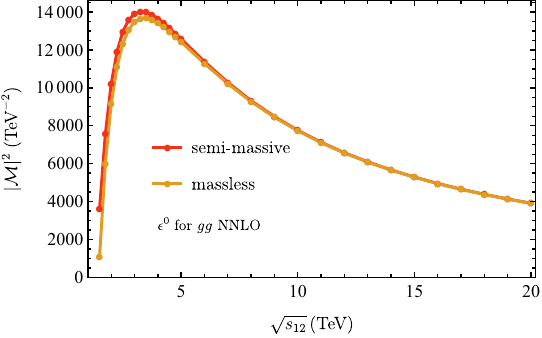}}
\subfigure{
\includegraphics[width=0.48\textwidth]{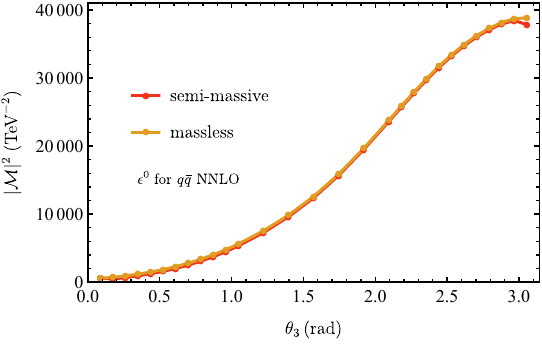}}
\subfigure{
\includegraphics[width=0.48\textwidth]{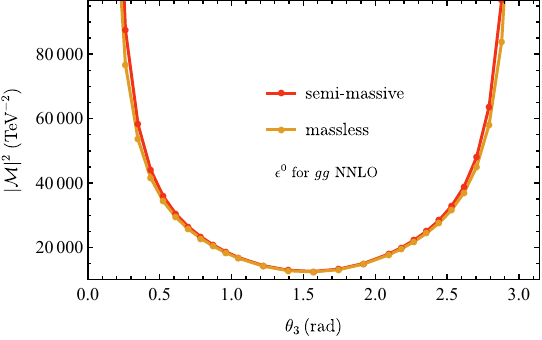}}
\subfigure{
\includegraphics[width=0.48\textwidth]{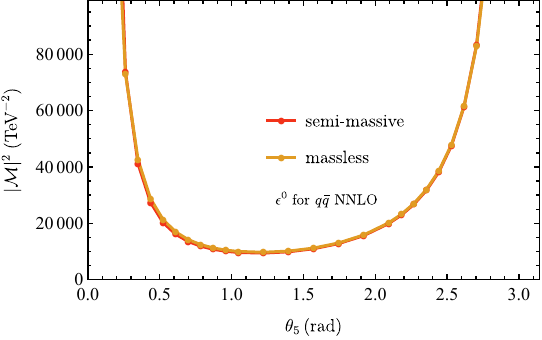}}
\subfigure{
\includegraphics[width=0.48\textwidth]{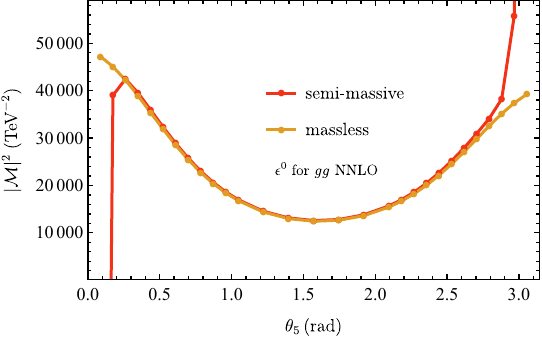}}
\caption{The finite parts of the NNLO squared amplitudes in the $q\bar{q}$ and $gg$ channels. The phase-space points are chosen as in Figs.~\ref{fig:diffscheme}, \ref{fig:diffschemetheta3s5} and \ref{fig:diffschemethetah5s5}.}
\label{fig:nnloeps0}
\end{figure}

We now present approximate predictions for the finite parts of the two-loop amplitudes for $t\bar{t}\gamma$ production. Figure~\ref{fig:nnloeps0} shows the results as functions of $\sqrt{s_{12}}$, $\theta_3$ and $\theta_5$, with phase-space points chosen as in Figs.~\ref{fig:diffscheme}, \ref{fig:diffschemetheta3s5} and \ref{fig:diffschemethetah5s5}. Since the exact results for these finite terms are unknown, we provide approximate results from both the massless and semi-massive schemes. The two schemes agree well at high energies but begin to deviate as $\sqrt{s_{12}}$ decreases. Based on our earlier findings, we conclude that the massless scheme yields a more reliable approximation at lower energies.

In addition to the squared amplitude as a whole, it is interesting to analyze its decomposition into color coefficients. The color structures of $t\bar{t}\gamma$ production are identical to those of $t\bar{t}$ and $t\bar{t}H$ production. We therefore adopt the color decomposition of Refs.~\cite{Czakon:2007ej, Czakon:2007wk, Ferroglia:2009ii,Wang:2024pmv}:
\begin{align}
\label{eq:color_dec}
   2\,{\rm Re} \Braket{{\cal M}_{q}^{(0)}|{\cal M}_{q}^{(2)}}
   &= 2(N^2-1)\,\bigg( N^2 A^q + B^q + \frac{1}{N^2}\,C^q 
    + N n_l\,D_l^q + N n_h\,D_h^q \nonumber
    \\
   &+ \frac{n_l}{N}\,E_l^q 
    + \frac{n_h}{N}\,E_h^q + n_l^2 F_l^q + n_l n_h\,F_{lh}^q 
    + n_h^2 F_h^q \bigg) \,, \nonumber
    \\
   2\,{\rm Re} \Braket{{\cal M}_{g}^{(0)}|{\cal M}_{g}^{(2)}}
   &= (N^2-1)\,\bigg( N^3 A^g + N\,B^g + \frac{1}{N}\,C^g 
    + \frac{1}{N^3}\,D^g \nonumber \\
   &+ N^2 n_l\,E_l^g + N^2 n_h\,E_h^g 
    + n_l\,F_l^g + n_h\,F_h^g + \frac{n_l}{N^2}\,G_l^g 
    + \frac{n_h}{N^2}\,G_h^g \nonumber\\
   &+ N n_l^2 H_l^g + N n_l n_h\,H_{lh}^g 
    + N n_h^2 H_h^g + \frac{n_l^2}{N}\,I_l^g 
    + \frac{n_l n_h}{N}\,I_{lh}^g + \frac{n_h^2}{N}\,I_h^g \bigg) \,.
\end{align}
In Tables~\ref{tab:qqnum1}, \ref{tab:ggnum1}, \ref{tab:qqnum2}, \ref{tab:ggnum2}, \ref{tab:qqnum3}, and \ref{tab:ggnum3}, we list the numerical values of the color-decomposed two-loop squared amplitudes at different orders in $\epsilon$. Results are shown at three representative phase-space points corresponding to relatively low, intermediate, and very high center-of-mass energies. We also perform a direct calculation of the $n_f^2$ terms at NNLO, which contribute to the coefficients $F^q$ in the $q\bar{q}$ channel and $H^g$, $I^g$ in the $gg$ channel. These results are also included in the tables. In the relatively low-energy region (Tables~\ref{tab:qqnum1} and~\ref{tab:ggnum1}), the relative differences between the massless-scheme results and the exact ones are significantly larger for $F^q_h$ ($H^g_h, I^g_h$) than for $F^q_l$ ($H^g_l, I^g_l$). In these cases, the closed quark loop effects factorize from the final-state top-quark pair, such that these coefficients share the same final-state contribution. Consequently, power corrections from the closed top-quark loop are more important than those from the external (anti-)top quark legs. This feature is interesting to explore further from the factorization perspective.\footnote{The NLP corrections from the closed top-quark loop in $F^q_h$ contribute entirely to the hard functions, while those in $H^g_h$ and $I^g_h$ contribute to both the hard functions and the gluon jet functions, according to the NLP factorization formula studied in Refs.~\cite{Laenen:2020nrt, vanBijleveld:2025ekz}.} A similar pattern is also observed for the terms proportional to $n_f$, namely the coefficients $D^q$ and $E^q$ in the $q\bar{q}$ channel, and $E^g$, $F^g$ and $G^g$ in the $gg$ channel. Overall, the massless scheme provides a reliable approximation to the exact results. These data may serve as a cross-check for a future exact evaluation of the two-loop squared amplitudes.

\section{Conclusion}\label{sec:conclusion}

In this paper, we have calculated the two-loop amplitudes for $t\bar{t}\gamma$ production at hadron colliders. First, for the infrared divergences, we employed the universal anomalous dimensions together with the one-loop massive amplitudes evaluated in dimensional regularization up to high order in $\epsilon$ (see Eq.~\eqref{eq:IRamps}). Second, for the finite parts, we used the mass-factorization formula in the high-energy boosted limit (see Eq.~\eqref{eq:QCDfac1}), where scalar products of external momenta are much larger than the top-quark mass. In this framework, the massive amplitudes are expressed in terms of the corresponding massless ones multiplied by universal factors that capture collinear and soft dynamics. To this end, we generated the two-loop massive amplitudes to include part of the power corrections beyond LP, and computed the scalar integrals with vanishing parton masses using IBP reduction together with existing results for the massless master integrals. The massive amplitudes were then constructed in two schemes: the massless scheme, where all partons are treated as massless, and the semi-massive scheme, where only the scalar Feynman integrals are taken to be massless. In addition, we calculated exactly the $n_f^2$ terms of the two-loop amplitudes.

We validated our approximations in the high-energy boosted limit from several perspectives. For the IR poles, we compared the predictions of both schemes with the exact results obtained in this work. For the finite remainders, we compared the approximate results with the exact ones for the $n_f^2$ terms. We found that the massless scheme provides a better approximation across most of phase space, and we confirmed that our approach correctly reproduces the scale dependence of the exact results. Finally, we presented predictions for the finite parts of the two-loop squared amplitudes, which—together with the two-loop IR poles—constitute the main new results of this work. We also provided results for the squared amplitudes at three representative phase-space points, including their decomposition into color coefficients.

We find that power corrections in the $q\bar{q}$ channel, especially in the relatively low-energy region, remain noticeable. Furthermore, the unknown power corrections from the scalar integrals are expected to significantly cancel the beyond-LP terms included in the semi-massive scheme. Improving the approximation therefore requires computing scalar integrals to higher powers in the high-energy limit, which can be achieved using strategies such as those described in Refs.~\cite{Davies:2018ood, Zhang:2024fcu, Guo:2025dlm}. Such results would also provide crucial input for studies of NLP factorization in the high-energy boosted limit, which we leave for future investigation. In addition, by combining our results with the real-emission contributions (double-real and real-virtual), one can compute the NNLO differential cross sections for $t\bar{t}\gamma$ production in the high-energy boosted limit.
Finally, by incorporating low-energy approximations such as threshold approximation and soft-photon approximation, and interpolating into the intermediate region, one would obtain a reasonable approximation across the whole phase space, enabling reliable predictions for differential cross sections from low to high energies.

\acknowledgments
  This work was supported in part by the National Natural Science Foundation of China under Grant No. 12375097, 12535003, 12347103, and the ERC grant No. 101041109 (``BOSON'').

\clearpage

\begin{table}[t!]
\begin{center}
\begin{tabular}{|c|r|r|r|r|r|}
    \hline
   \text{exact}&\multirow{2}*{$\eps^{-4}$}& \multirow{2}*{$\eps^{-3}$}&\multirow{2}*{$\eps^{-2}$}&\multirow{2}*{$\eps^{-1}$}&\multirow{2}*{$\eps^{0}$}\\ \cline{1-1}
   \text{massless}&  &  &  &  & \\ \hline
   
\multirow{2}*{$A^q$}  &   $0.2283598$  &   $-1.589481$  &   $8.515970$  &   $-43.60179$  &   \\ \cline{2-6}   &   $0.2090052$  &   $-1.479696$  &   $7.815084$  &   $-39.32127$  &   $154.5011$\\ \hline
 \multirow{2}*{$B^q$}  &   $-0.4567197$  &   $4.847678$  &   $-21.98068$  &   $19.67059$  &   \\ \cline{2-6}   &   $-0.4180103$  &   $4.491915$  &   $-20.67344$  &   $62.01943$  &   $-148.0833$\\ \hline
 \multirow{2}*{$C^q$}  &   $0.2283598$  &   $-3.258197$  &   $5.382477$  &   $33.15795$  &   \\ \cline{2-6}   &   $0.2090052$  &   $-3.012218$  &   $5.484770$  &   $34.93798$  &   $-160.1622$\\ \hline
 \multirow{2}*{$D_l^q$}  &   $0$  &   $-0.2283598$  &   $0.7189743$  &   $1.638202$  &   \\ \cline{2-6}   &   $0$  &   $-0.2090052$  &   $0.6956674$  &   $1.383895$  &   $-13.45819$\\ \hline
 \multirow{2}*{$D_h^q$}  &   $0$  &   $0$  &   $-0.6806559$  &   $4.903448$  &   \\ \cline{2-6}   &   $0$  &   $0$  &   $-0.6527637$  &   $4.843065$  &   $-15.46237$\\ \hline
 \multirow{2}*{$E_l^q$}  &   $0$  &   $0.2283598$  &   $-0.8565532$  &   $4.282602$  &   \\ \cline{2-6}   &   $0$  &   $0.2090052$  &   $-0.8233320$  &   $4.057688$  &   $-33.89710$\\ \hline
 \multirow{2}*{$E_h^q$}  &   $0$  &   $0$  &   $0.6806559$  &   $0.4796641$  &   \\ \cline{2-6}   &   $0$  &   $0$  &   $0.6527637$  &   $0.05964861$  &   $-30.60037$\\ \hline
 \multirow{2}*{$F_l^q$}  &   $0$  &   $0$  &   $0$  &   $0$  &  $-0.8284431$ \\ \cline{2-6}   &   $0$  &   $0$  &   $0$  &   $0$  &   $-0.7701217$\\ \hline
 \multirow{2}*{$F_{lh}^q$}  &   $0$  &   $0$  &   $0$  &   $0$  &  $-1.760584$ \\ \cline{2-6}   &   $0$  &   $0$  &   $0$  &   $0$  &   $-1.540243$\\ \hline
 \multirow{2}*{$F_h^q$}  &   $0$  &   $0$  &   $0$  &   $0$  &  $-0.9285263$ \\ \cline{2-6}   &   $0$  &   $0$  &   $0$  &   $0$  &   $-0.7701217$\\ \hline
 \multirow{2}*{$\text{Total}$}  &   $1.623892$  &   $-12.86447$  &   $62.80305$  &   $-322.4803$  &   \\ \cline{2-6}   &   $1.486259$  &   $-11.94678$  &   $57.59382$  &   $-245.9197$  &   $881.9512$\\ \hline
 $\text{Ratio}$  &   0.9152  &   0.9287  &   0.9171  &   0.7626  &    \\ \hline

\end{tabular}
\caption{\label{tab:qqnum1} Color-decomposed NNLO squared amplitude in the $q\bar{q}$ channel at the phase-space point $\sqrt{s_{12}} = \SI{2}{TeV}$, $m_t=\SI{173}{GeV}$, $\theta_3 = 14\pi/29$, $\phi_3=34\pi/29$, $\theta_5=15\pi/29$ and $q_5=20\,q_{5,\rm max}/29$, which corresponds to $(\sqrt{|s_{13}|}, \sqrt{|s_{14}|}, \sqrt{|s_{23}|}, \sqrt{|s_{24}|}) = (1.11398, 1.14698, 1.17601, 1.14968)$~TeV.}
\end{center}
\end{table}

\begin{table}[t!]
\begin{center}
\begin{tabular}{|c|r|r|r|r|r|}
    \hline
   \text{exact}&\multirow{2}*{$\eps^{-4}$}& \multirow{2}*{$\eps^{-3}$}&\multirow{2}*{$\eps^{-2}$}&\multirow{2}*{$\eps^{-1}$}&\multirow{2}*{$\eps^{0}$}\\ \cline{1-1}
   \text{massless}&  &  &  &  & \\ \hline   

\multirow{2}*{$A^g$}  &   $0.2109546$  &   $-1.988709$  &   $9.022201$  &   $-28.20074$  &   \\ \cline{2-6}   &   $0.1841424$  &   $-1.844645$  &   $8.279166$  &   $-26.21126$  &   $50.66531$\\ \hline
 \multirow{2}*{$B^g$}  &   $-0.3135349$  &   $3.205230$  &   $-18.92590$  &   $101.0938$  &   \\ \cline{2-6}   &   $-0.2685287$  &   $2.907657$  &   $-16.65607$  &   $79.27004$  &   $-285.3764$\\ \hline
 \multirow{2}*{$C^g$}  &   $0$  &   $-0.8612544$  &   $7.135878$  &   $-7.013240$  &   \\ \cline{2-6}   &   $0$  &   $-0.7360521$  &   $6.007253$  &   $-18.09636$  &   $21.29864$\\ \hline
 \multirow{2}*{$D^g$}  &   $0$  &   $0$  &   $0.1839062$  &   $-4.209482$  &   \\ \cline{2-6}   &   $0$  &   $0$  &   $0.1581783$  &   $-4.218801$  &   $18.13444$\\ \hline
 \multirow{2}*{$E_l^g$}  &   $0$  &   $-0.2461137$  &   $1.996011$  &   $-7.312272$  &   \\ \cline{2-6}   &   $0$  &   $-0.2148328$  &   $1.882865$  &   $-6.965950$  &   $18.17611$\\ \hline
 \multirow{2}*{$E_h^g$}  &   $0$  &   $0$  &   $0.2369820$  &   $-2.313407$  &   \\ \cline{2-6}   &   $0$  &   $0$  &   $0.1577332$  &   $-1.802488$  &   $7.541058$\\ \hline
 \multirow{2}*{$F_l^g$}  &   $0$  &   $0.3657907$  &   $-3.556003$  &   $16.70099$  &   \\ \cline{2-6}   &   $0$  &   $0.3132835$  &   $-3.298786$  &   $15.83586$  &   $-49.45538$\\ \hline
 \multirow{2}*{$F_h^g$}  &   $0$  &   $0$  &   $-0.9923075$  &   $10.45595$  &   \\ \cline{2-6}   &   $0$  &   $0$  &   $-0.6742469$  &   $8.237854$  &   $-36.20808$\\ \hline
 \multirow{2}*{$G_l^g$}  &   $0$  &   $0$  &   $0.4306272$  &   $-3.037363$  &   \\ \cline{2-6}   &   $0$  &   $0$  &   $0.3680260$  &   $-2.794666$  &   $0.02136906$\\ \hline
 \multirow{2}*{$G_h^g$}  &   $0$  &   $0$  &   $0$  &   $-1.635062$  &   \\ \cline{2-6}   &   $0$  &   $0$  &   $0$  &   $-1.049551$  &   $-0.5911322$\\ \hline
 \multirow{2}*{$H_l^g$}  &   $0$  &   $0$  &   $0.04687880$  &   $-0.2606374$  &  $0.4432391$ \\ \cline{2-6}   &   $0$  &   $0$  &   $0.04092053$  &   $-0.2562426$  &   $0.4291040$\\ \hline
 \multirow{2}*{$H_{lh}^g$}  &   $0$  &   $0$  &   $0$  &   $-0.07899399$  &  $0.1054877$ \\ \cline{2-6}   &   $0$  &   $0$  &   $0$  &   $-0.05257773$  &   $0.1339042$\\ \hline
 \multirow{2}*{$H_h^g$}  &   $0$  &   $0$  &   $0$  &   $0$  & $-0.2011094$  \\ \cline{2-6}   &   $0$  &   $0$  &   $0$  &   $0$  &   $-0.1258135$\\ \hline
 \multirow{2}*{$I_l^g$}  &   $0$  &   $0$  &   $-0.06967442$  &   $0.5422631$  & $-1.501144$  \\ \cline{2-6}   &   $0$  &   $0$  &   $-0.05967305$  &   $0.5217467$  &   $-1.443137$\\ \hline
 \multirow{2}*{$I_{lh}^g$}  &   $0$  &   $0$  &   $0$  &   $0.3307692$  &  $-0.5663351$ \\ \cline{2-6}   &   $0$  &   $0$  &   $0$  &   $0.2247490$  &   $-0.6462586$\\ \hline
 \multirow{2}*{$I_h^g$}  &   $0$  &   $0$  &   $0$  &   $0$  &  $0.8335352$ \\ \cline{2-6}   &   $0$  &   $0$  &   $0$  &   $0$  &   $0.5498681$\\ \hline
 \multirow{2}*{$\text{Total}$}  &   $4.755170$  &   $-53.61269$  &   $265.5627$  &   $-734.0757$  &   \\ \cline{2-6}   &   $4.166258$  &   $-49.42885$  &   $247.3341$  &   $-735.3090$  &   $1142.755$\\ \hline
 $\text{Ratio}$  &   0.8762  &   0.9220  &   0.9314  &   1.002  &    \\ \hline

\end{tabular}
\caption{\label{tab:ggnum1} Color-decomposed NNLO squared amplitude in the $gg$ channel at the phase-space point $\sqrt{s_{12}} = \SI{2}{TeV}$, $m_t=\SI{173}{GeV}$, $\theta_3 = 14\pi/29$, $\phi_3=34\pi/29$, $\theta_5=15\pi/29$ and $q_5=20\,q_{5,\rm max}/29$, which corresponds to $(\sqrt{|s_{13}|}, \sqrt{|s_{14}|}, \sqrt{|s_{23}|}, \sqrt{|s_{24}|}) = (1.11398, 1.14698, 1.17601, 1.14968)$~TeV.}
\end{center}
\end{table}

\begin{table}[t!]
\begin{center}
\begin{tabular}{|c|r|r|r|r|r|}
    \hline
   \text{exact}&\multirow{2}*{$\eps^{-4}$}& \multirow{2}*{$\eps^{-3}$}&\multirow{2}*{$\eps^{-2}$}&\multirow{2}*{$\eps^{-1}$}&\multirow{2}*{$\eps^{0}$}\\ \cline{1-1}
   \text{massless}&  &  &  &  & \\ \hline

\multirow{2}*{$A^q$}  &   $0.05264489$  &   $-0.8045300$  &   $7.571800$  &   $-50.38964$  &   \\ \cline{2-6}   &   $0.05206390$  &   $-0.7965400$  &   $7.501998$  &   $-49.94393$  &   $228.2212$\\ \hline
 \multirow{2}*{$B^q$}  &   $-0.1052898$  &   $2.258002$  &   $-20.75968$  &   $99.96689$  &   \\ \cline{2-6}   &   $-0.1041278$  &   $2.236790$  &   $-20.63005$  &   $118.4241$  &   $-459.4837$\\ \hline
 \multirow{2}*{$C^q$}  &   $0.05264489$  &   $-1.453472$  &   $12.38175$  &   $-49.99110$  &   \\ \cline{2-6}   &   $0.05206390$  &   $-1.440250$  &   $12.34102$  &   $-50.17464$  &   $121.6998$\\ \hline
 \multirow{2}*{$D_l^q$}  &   $0$  &   $-0.05264489$  &   $0.1696066$  &   $2.503278$  &   \\ \cline{2-6}   &   $0$  &   $-0.05206390$  &   $0.1689226$  &   $2.470040$  &   $-22.47169$\\ \hline
 \multirow{2}*{$D_h^q$}  &   $0$  &   $0$  &   $-0.2973506$  &   $3.817902$  &   \\ \cline{2-6}   &   $0$  &   $0$  &   $-0.2951209$  &   $3.796775$  &   $-24.12762$\\ \hline
 \multirow{2}*{$E_l^q$}  &   $0$  &   $0.05264489$  &   $-0.2894049$  &   $-1.683770$  &   \\ \cline{2-6}   &   $0$  &   $0.05206390$  &   $-0.2880421$  &   $-1.659577$  &   $7.858607$\\ \hline
 \multirow{2}*{$E_h^q$}  &   $0$  &   $0$  &   $0.2973506$  &   $-3.447385$  &   \\ \cline{2-6}   &   $0$  &   $0$  &   $0.2951209$  &   $-3.439339$  &   $10.59126$\\ \hline
 \multirow{2}*{$F_l^q$}  &   $0$  &   $0$  &   $0$  &   $0$  &  $0.3985754$ \\ \cline{2-6}   &   $0$  &   $0$  &   $0$  &   $0$  &   $0.3930888$\\ \hline
 \multirow{2}*{$F_{lh}^q$}  &   $0$  &   $0$  &   $0$  &   $0$  &  $0.7902131$ \\ \cline{2-6}   &   $0$  &   $0$  &   $0$  &   $0$  &   $0.7861776$\\ \hline
 \multirow{2}*{$F_h^q$}  &   $0$  &   $0$  &   $0$  &   $0$  &  $0.3916563$ \\ \cline{2-6}   &   $0$  &   $0$  &   $0$  &   $0$  &   $0.3930888$\\ \hline
 \multirow{2}*{$\text{Total}$}  &   $0.3743637$  &   $-5.846197$  &   $50.03109$  &   $-314.0470$  &   \\ \cline{2-6}   &   $0.3702322$  &   $-5.786283$  &   $49.52594$  &   $-292.1177$  &   $1229.350$\\ \hline
 $\text{Ratio}$  &   0.9890  &   0.9898  &   0.9899  &   0.9302  &    \\ \hline

\end{tabular}
\caption{\label{tab:qqnum2} Color-decomposed NNLO squared amplitude in the $q\bar{q}$ channel at the phase-space point $\sqrt{s_{12}} = \SI{5}{TeV}$, $m_t=\SI{173}{GeV}$, $\theta_3 = 11\pi/18$, $\phi_3=34\pi/29$, $\theta_5=15\pi/29$ and $q_5=20\,q_{5,\rm max}/29$, which corresponds to $(\sqrt{|s_{13}|}, \sqrt{|s_{14}|}, \sqrt{|s_{23}|}, \sqrt{|s_{24}|}) = (3.24608, 2.32073, 2.27293, 3.41892)$~TeV.}
\end{center}
\end{table}

\begin{table}[t!]
\begin{center}
\begin{tabular}{|c|r|r|r|r|r|}
    \hline
   \text{exact}&\multirow{2}*{$\eps^{-4}$}& \multirow{2}*{$\eps^{-3}$}&\multirow{2}*{$\eps^{-2}$}&\multirow{2}*{$\eps^{-1}$}&\multirow{2}*{$\eps^{0}$}\\ \cline{1-1}
   \text{massless}&  &  &  &  & \\ \hline

\multirow{2}*{$A^g$}  &   $0.04420994$  &   $-0.6366716$  &   $4.728640$  &   $-23.69891$  &   \\ \cline{2-6}   &   $0.04306350$  &   $-0.6249713$  &   $4.641818$  &   $-23.29684$  &   $80.77025$\\ \hline
 \multirow{2}*{$B^g$}  &   $-0.05722378$  &   $0.9839884$  &   $-8.537617$  &   $50.90050$  &   \\ \cline{2-6}   &   $-0.05545865$  &   $0.9607114$  &   $-8.321031$  &   $48.14527$  &   $-199.2297$\\ \hline
 \multirow{2}*{$C^g$}  &   $0$  &   $-0.2609864$  &   $3.110131$  &   $-11.47391$  &   \\ \cline{2-6}   &   $0$  &   $-0.2529048$  &   $3.003469$  &   $-15.86476$  &   $50.67867$\\ \hline
 \multirow{2}*{$D^g$}  &   $0$  &   $0$  &   $-0.1563746$  &   $0.1825180$  &   \\ \cline{2-6}   &   $0$  &   $0$  &   $-0.1514881$  &   $0.1228716$  &   $3.884178$\\ \hline
 \multirow{2}*{$E_l^g$}  &   $0$  &   $-0.05157826$  &   $0.5237598$  &   $-2.433499$  &   \\ \cline{2-6}   &   $0$  &   $-0.05024075$  &   $0.5163243$  &   $-2.405636$  &   $7.614495$\\ \hline
 \multirow{2}*{$E_h^g$}  &   $0$  &   $0$  &   $0.03150154$  &   $-0.4511388$  &   \\ \cline{2-6}   &   $0$  &   $0$  &   $0.02738597$  &   $-0.4066532$  &   $2.474837$\\ \hline
 \multirow{2}*{$F_l^g$}  &   $0$  &   $0.06676107$  &   $-0.8408940$  &   $4.923497$  &   \\ \cline{2-6}   &   $0$  &   $0.06470175$  &   $-0.8251418$  &   $4.853851$  &   $-20.01351$\\ \hline
 \multirow{2}*{$F_h^g$}  &   $0$  &   $0$  &   $-0.1344545$  &   $2.108986$  &   \\ \cline{2-6}   &   $0$  &   $0$  &   $-0.1175444$  &   $1.915518$  &   $-13.11469$\\ \hline
 \multirow{2}*{$G_l^g$}  &   $0$  &   $0$  &   $0.1304932$  &   $-1.031368$  &   \\ \cline{2-6}   &   $0$  &   $0$  &   $0.1264524$  &   $-1.009265$  &   $2.874627$\\ \hline
 \multirow{2}*{$G_h^g$}  &   $0$  &   $0$  &   $0$  &   $-0.3601850$  &   \\ \cline{2-6}   &   $0$  &   $0$  &   $0$  &   $-0.3153018$  &   $1.789123$\\ \hline
 \multirow{2}*{$H_l^g$}  &   $0$  &   $0$  &   $0.009824431$  &   $-0.05312400$  & $0.06954816$  \\ \cline{2-6}   &   $0$  &   $0$  &   $0.009569668$  &   $-0.05299359$  &   $0.06899445$\\ \hline
 \multirow{2}*{$H_{lh}^g$}  &   $0$  &   $0$  &   $0$  &   $-0.01050051$  &  $-0.01427645$ \\ \cline{2-6}   &   $0$  &   $0$  &   $0$  &   $-0.009128657$  &   $-0.01275757$\\ \hline
 \multirow{2}*{$H_h^g$}  &   $0$  &   $0$  &   $0$  &   $0$  &  $-0.04904650$ \\ \cline{2-6}   &   $0$  &   $0$  &   $0$  &   $0$  &   $-0.04264576$\\ \hline
 \multirow{2}*{$I_l^g$}  &   $0$  &   $0$  &   $-0.01271639$  &   $0.09643057$  & $-0.2131644$  \\ \cline{2-6}   &   $0$  &   $0$  &   $-0.01232414$  &   $0.09567221$  &   $-0.2107835$\\ \hline
 \multirow{2}*{$I_{lh}^g$}  &   $0$  &   $0$  &   $0$  &   $0.04481815$  &  $0.02864711$ \\ \cline{2-6}   &   $0$  &   $0$  &   $0$  &   $0.03918146$  &   $0.02365909$\\ \hline
 \multirow{2}*{$I_h^g$}  &   $0$  &   $0$  &   $0$  &   $0$  &  $0.2095268$ \\ \cline{2-6}   &   $0$  &   $0$  &   $0$  &   $0$  &   $0.1840802$\\ \hline
 \multirow{2}*{$\text{Total}$}  &   $1.021997$  &   $-16.31238$  &   $123.3085$  &   $-581.7046$  &   \\ \cline{2-6}   &   $0.9963387$  &   $-16.01372$  &   $121.2846$  &   $-579.4359$  &   $1856.883$\\ \hline
 $\text{Ratio}$  &   0.9749  &   0.9817  &   0.9836  &   0.9961  &    \\ \hline

\end{tabular}
\caption{\label{tab:ggnum2} Color-decomposed NNLO squared amplitude in the $gg$ channel at the phase-space point $\sqrt{s_{12}} = \SI{5}{TeV}$, $m_t=\SI{173}{GeV}$, $\theta_3 = 11\pi/18$, $\phi_3=34\pi/29$, $\theta_5=15\pi/29$ and $q_5=20\,q_{5,\rm max}/29$, which corresponds to $(\sqrt{|s_{13}|}, \sqrt{|s_{14}|}, \sqrt{|s_{23}|}, \sqrt{|s_{24}|}) = (3.24608, 2.32073, 2.27293, 3.41892)$~TeV.}
\end{center}
\end{table}

\begin{table}[t!]
\begin{center}
\begin{tabular}{|c|r|r|r|r|r|}
    \hline
   \text{exact}&\multirow{2}*{$\eps^{-4}$}& \multirow{2}*{$\eps^{-3}$}&\multirow{2}*{$\eps^{-2}$}&\multirow{2}*{$\eps^{-1}$}&\multirow{2}*{$\eps^{0}$}\\ \cline{1-1}
   \text{massless}&  &  &  &  & \\ \hline
   
\multirow{2}*{$A^q$}  &   $0.008313164$  &   $-0.1658354$  &   $1.877353$  &   $-14.38264$  &   \\ \cline{2-6}   &   $0.008282092$  &   $-0.1652614$  &   $1.871562$  &   $-14.34242$  &   $75.08431$\\ \hline
 \multirow{2}*{$B^q$}  &   $-0.01662633$  &   $0.3925240$  &   $-4.099684$  &   $21.74389$  &   \\ \cline{2-6}   &   $-0.01656418$  &   $0.3911591$  &   $-4.089273$  &   $25.60180$  &   $-106.5415$\\ \hline
 \multirow{2}*{$C^q$}  &   $0.008313164$  &   $-0.2266886$  &   $1.928764$  &   $-6.741097$  &   \\ \cline{2-6}   &   $0.008282092$  &   $-0.2258977$  &   $1.925272$  &   $-6.742960$  &   $7.790823$\\ \hline
 \multirow{2}*{$D_l^q$}  &   $0$  &   $-0.008313164$  &   $0.02783290$  &   $0.7088284$  &   \\ \cline{2-6}   &   $0$  &   $-0.008282092$  &   $0.02779463$  &   $0.7058663$  &   $-7.464632$\\ \hline
 \multirow{2}*{$D_h^q$}  &   $0$  &   $0$  &   $-0.06143042$  &   $0.9962927$  &   \\ \cline{2-6}   &   $0$  &   $0$  &   $-0.06124643$  &   $0.9938256$  &   $-7.906412$\\ \hline
 \multirow{2}*{$E_l^q$}  &   $0$  &   $0.008313164$  &   $-0.03287652$  &   $-0.5246784$  &   \\ \cline{2-6}   &   $0$  &   $0.008282092$  &   $-0.03282291$  &   $-0.5223829$  &   $2.575036$\\ \hline
 \multirow{2}*{$E_h^q$}  &   $0$  &   $0$  &   $0.06143042$  &   $-0.8334049$  &   \\ \cline{2-6}   &   $0$  &   $0$  &   $0.06124643$  &   $-0.8316262$  &   $3.067942$\\ \hline
 \multirow{2}*{$F_l^q$}  &   $0$  &   $0$  &   $0$  &   $0$  &  $0.1562176$ \\ \cline{2-6}   &   $0$  &   $0$  &   $0$  &   $0$  &   $0.1555812$\\ \hline
 \multirow{2}*{$F_{lh}^q$}  &   $0$  &   $0$  &   $0$  &   $0$  &  $0.3120468$ \\ \cline{2-6}   &   $0$  &   $0$  &   $0$  &   $0$  &   $0.3111624$\\ \hline
 \multirow{2}*{$F_h^q$}  &   $0$  &   $0$  &   $0$  &   $0$  &  $0.1558294$ \\ \cline{2-6}   &   $0$  &   $0$  &   $0$  &   $0$  &   $0.1555812$\\ \hline
 \multirow{2}*{$\text{Total}$}  &   $0.05911583$  &   $-1.236024$  &   $13.20969$  &   $-95.97980$  &   \\ \cline{2-6}   &   $0.05889488$  &   $-1.231721$  &   $13.16760$  &   $-91.80754$  &   $445.3095$\\ \hline
 $\text{Ratio}$  &   0.9963  &   0.9965  &   0.9968  &   0.9565  &    \\ \hline

\end{tabular}
\caption{Color-decomposed NNLO squared amplitude in the $q\bar{q}$ channel at the phase-space point $\sqrt{s_{12}} = \SI{10}{TeV}$, $m_t=\SI{173}{GeV}$, $\theta_3 = 14\pi/29$, $\phi_3=34\pi/29$, $\theta_5=15\pi/29$ and $q_5=20\,q_{5,\rm max}/29$, which corresponds to $( \sqrt{|s_{13}|}, \sqrt{|s_{14}|}, \sqrt{|s_{23}|}, \sqrt{|s_{24}|}) = (5.54848, 5.73368, 5.85746, 5.75183)$~TeV.}
\label{tab:qqnum3}
\end{center}
\end{table}

\begin{table}[t!]
\begin{center}
\begin{tabular}{|l|r|r|r|r|r|}
    \hline
   \text{exact}&\multirow{2}*{$\eps^{-4}$}& \multirow{2}*{$\eps^{-3}$}&\multirow{2}*{$\eps^{-2}$}&\multirow{2}*{$\eps^{-1}$}&\multirow{2}*{$\eps^{0}$}\\ \cline{1-1}
   \text{semi-}&  &  &  &  & \\ \hline
   
\multirow{2}*{$A^g$}  &   $0.007372706$  &   $-0.1447482$  &   $1.440026$  &   $-9.412260$  &   \\ \cline{2-6}   &   $0.007327841$  &   $-0.1440745$  &   $1.433824$  &   $-9.375672$  &   $42.71008$\\ \hline
 \multirow{2}*{$B^g$}  &   $-0.01067604$  &   $0.2421086$  &   $-2.668968$  &   $19.32902$  &   \\ \cline{2-6}   &   $-0.01060172$  &   $0.2407542$  &   $-2.653716$  &   $18.81769$  &   $-93.79331$\\ \hline
 \multirow{2}*{$C^g$}  &   $0$  &   $-0.06346660$  &   $0.9820515$  &   $-5.639019$  &   \\ \cline{2-6}   &   $0$  &   $-0.06302361$  &   $0.9743994$  &   $-6.825529$  &   $29.21020$\\ \hline
 \multirow{2}*{$D^g$}  &   $0$  &   $0$  &   $-0.06798140$  &   $0.3848235$  &   \\ \cline{2-6}   &   $0$  &   $0$  &   $-0.06750472$  &   $0.3764003$  &   $-0.06792920$\\ \hline
 \multirow{2}*{$E_l^g$}  &   $0$  &   $-0.008601490$  &   $0.1108373$  &   $-0.6520855$  &   \\ \cline{2-6}   &   $0$  &   $-0.008549148$  &   $0.1104304$  &   $-0.6503525$  &   $2.590406$\\ \hline
 \multirow{2}*{$E_h^g$}  &   $0$  &   $0$  &   $0.006870839$  &   $-0.1313022$  &   \\ \cline{2-6}   &   $0$  &   $0$  &   $0.006473316$  &   $-0.1256742$  &   $0.9880121$\\ \hline
 \multirow{2}*{$F_l^g$}  &   $0$  &   $0.01245538$  &   $-0.1950641$  &   $1.418727$  &   \\ \cline{2-6}   &   $0$  &   $0.01236868$  &   $-0.1942021$  &   $1.414256$  &   $-7.417823$\\ \hline
 \multirow{2}*{$F_h^g$}  &   $0$  &   $0$  &   $-0.02928135$  &   $0.5932396$  &   \\ \cline{2-6}   &   $0$  &   $0$  &   $-0.02764421$  &   $0.5688947$  &   $-5.029673$\\ \hline
 \multirow{2}*{$G_l^g$}  &   $0$  &   $0$  &   $0.03173330$  &   $-0.2959568$  &   \\ \cline{2-6}   &   $0$  &   $0$  &   $0.03151180$  &   $-0.2944347$  &   $1.195346$\\ \hline
 \multirow{2}*{$G_h^g$}  &   $0$  &   $0$  &   $0$  &   $-0.09396758$  &   \\ \cline{2-6}   &   $0$  &   $0$  &   $0$  &   $-0.08833384$  &   $0.7660025$\\ \hline
 \multirow{2}*{$H_l^g$}  &   $0$  &   $0$  &   $0.001638379$  &   $-0.01026111$  & $0.01707854$  \\ \cline{2-6}   &   $0$  &   $0$  &   $0.001628409$  &   $-0.01025332$  &   $0.01705303$\\ \hline
 \multirow{2}*{$H_{lh}^g$}  &   $0$  &   $0$  &   $0$  &   $-0.002290280$  & $-0.002135886$ \\  \cline{2-6}   &   $0$  &   $0$  &   $0$  &   $-0.002157772$  &   $-0.001931145$\\ \hline
 \multirow{2}*{$H_h^g$}  &   $0$  &   $0$  &   $0$  &   $0$   &  $-0.01307502$ \\ \cline{2-6}   &   $0$  &   $0$  &   $0$  &   $0$  &   $-0.01224576$\\ \hline
 \multirow{2}*{$I_l^g$}  &   $0$  &   $0$  &   $-0.002372453$  &   $0.02096296$  &  $-0.05739759$ \\ \cline{2-6}   &   $0$  &   $0$  &   $-0.002355939$  &   $0.02092716$  &   $-0.05729325$\\ \hline
 \multirow{2}*{$I_{lh}^g$}  &   $0$  &   $0$  &   $0$  &   $0.009760449$  &  $0.006012662$ \\ \cline{2-6}   &   $0$  &   $0$  &   $0$  &   $0.009214735$  &   $0.005201770$\\ \hline
 \multirow{2}*{$I_h^g$}  &   $0$  &   $0$  &   $0$  &   $0$  &  $0.05610000$ \\ \cline{2-6}   &   $0$  &   $0$  &   $0$  &   $0$  &   $0.05274607$\\ \hline
 \multirow{2}*{$\text{Total}$}  &   $0.1670349$  &   $-3.527821$  &   $35.36427$  &   $-221.6359$  &   \\ \cline{2-6}   &   $0.1660465$  &   $-3.511624$  &   $35.22337$  &   $-222.4931$  &   $966.3794$\\ \hline
 $\text{Ratio}$  &   0.9941  &   0.9954  &   0.9960  &   1.004  &    \\ \hline

\end{tabular}
\caption{Color-decomposed NNLO squared amplitude in the $gg$ channel at the phase-space point $\sqrt{s_{12}} = \SI{10}{TeV}$, $m_t=\SI{173}{GeV}$, $\theta_3 = 14\pi/29$, $\phi_3=34\pi/29$, $\theta_5=15\pi/29$ and $q_5=20\,q_{5,\rm max}/29$, which corresponds to $( \sqrt{|s_{13}|}, \sqrt{|s_{14}|}, \sqrt{|s_{23}|}, \sqrt{|s_{24}|}) = (5.54848, 5.73368, 5.85746, 5.75183)$~TeV.}
\label{tab:ggnum3}
\end{center}
\end{table}

\clearpage

\appendix

\section{Renormalization constant}\label{sec:renormalconstant}
In this section, we present the renormalization constants needed in our work. Up to NNLO, the renormalization constant of QCD coupling constant $\alpha_s$ is given by 
\begin{equation}
Z_{\alpha_s}=1-\lt(\fc{\alpha_s}{4\pi}\rt)\fc{\beta_0}{\eps}+\lt(\fc{\alpha_s}{4\pi}\rt)^2\lt(\fc{\beta_0^2}{\eps^2}-\fc{\beta_1}{2\eps}\rt) \,,
\end{equation}
where
\begin{equation}
\beta_0=\fc{11}{3}C_A-\fc{4}{3}T_Fn_f,\quad \beta_1=\fc{34}{3}C_A^2-\fc{20}{3}C_AT_Fn_f-4C_FT_Fn_f.
\end{equation}
Up to NNLO, the top-quark mass renormalization constant in the on-shell scheme is given by:
\begin{align}
Z_m &= 1 + \frac{\alpha_s}{4\pi} C_F \left[ -\frac{3}{\epsilon} - \left(4+3L_\mu\right) - \epsilon\left(8+4L_\mu+\frac{\pi^2}{4}+\frac{3}{2}L_\mu^2\right) \right. \nonumber \\ 
& \left.+\epsilon^2\left(-L_{\mu}(8+\frac{\pi^2}{4}) -2L_{\mu}^2-\frac{L_{\mu}^3}{2} -48-\frac{\pi^2}{3}+\zeta_3 \right)\right]  \nonumber \\
&+ \left(\frac{\alpha_s}{4\pi}\right)^2 \bigg\{C_A \left[\frac{11}{2 \epsilon ^2}-\frac{97}{12 \epsilon
   }+\frac{4 \pi ^2}{3}-\frac{1111}{24}+6 \zeta_3-\frac{11 L_{\mu }^2}{2} -4 \pi ^2 L_{\mu }-\frac{185 L_{\mu }}{6}\right]\nonumber \\
&+C_F \left[\frac{9}{2 \epsilon ^2}+\frac{36 L_{\mu
   }+45}{4\epsilon } +\frac{1}{8} \left(-96 \zeta_3+199-34 \pi ^2\right) + 9 L_{\mu }^2+8 \pi ^2 L_{\mu } +\frac{45 L_{\mu }}{2}\right]\nonumber \\
& + T_Fn_h
   \bigg(-\frac{2}{\epsilon ^2}+\frac{5}{3 \epsilon } +2 L_{\mu }^2+\frac{26 L_{\mu }}{3}-\frac{8 \pi ^2}{3}+\frac{143}{6}\bigg)\nonumber \\
&+T_Fn_l
   \bigg[-\frac{2}{\epsilon ^2}+\frac{5}{3 \epsilon } +\frac{1}{6} \left(71+8 \pi ^2\right)+2 L_{\mu }^2+\frac{26 L_{\mu }}{3}\bigg]\bigg\}+ \mathcal{O}(\alpha_s^3) \, ,
\end{align}
where $L_{\mu} = \ln\left(\mu^2/m_t^2\right)$. The on-shell wave-function renormalization constants are
\begin{align}
Z_{q} &= 1 + \left(\frac{\alpha_s}{4\pi}\right)^2 C_F \left(\frac{1}{2\epsilon} -\frac{5}{12} + L_{\mu}\right) + \mathcal{O}(\alpha_s^3) \, , \nonumber
\\
Z_{g} &= 1 + \frac{\alpha_s}{4\pi} T_Fn_h \left[ -\frac{4}{3\epsilon} - \frac{4}{3}L_{\mu} - \epsilon\left(\frac{\pi^2}{9} + \frac{2}{3}L^2_{\mu} \right)+\epsilon^2\left( -\frac{\pi^2}{9} L_{\mu}-\frac{2}{9} L_{\mu}^3 +\frac{4}{9}\zeta_3\right)\right] \nonumber \\ 
&+ \left(\frac{\alpha_s}{4\pi}\right)^2 T_Fn_h \left\{C_A \left[\frac{35}{9 \epsilon
   ^2}-\frac{1}{\epsilon}\left(\frac{5}{2} - \frac{26 L_{\mu }}{9}\right)+\frac{13}{12}-5 L_{\mu }+\frac{13 \pi ^2}{54}+\frac{4 L_{\mu }^2}{9}\right] \right. \nonumber \\
&\left.-C_F \left(\frac{2}{\epsilon }+15+4 L_{\mu}\right)+n_h \left(\frac{8 L_{\mu }}{9 \epsilon}+\frac{2 \pi ^2}{27}+ \frac{4 L_{\mu }^2}{3}\right) \right. \nonumber \\
&\left.-n_l \left(\frac{8}{9 \epsilon ^2}+\frac{8 L_{\mu }}{9 \epsilon}+\frac{2 \pi ^2}{27}+\frac{4 L_{\mu }^2}{9}\right)\right\}+ \mathcal{O}(\alpha_s^3) \, , \nonumber
\\
Z_{Q} &= 1 + \frac{\alpha_s}{4\pi} C_F \left[ -\frac{3}{\epsilon} - \left(4+3L_\mu\right) - \epsilon\left(8+4L_\mu+\frac{\pi^2}{4}+\frac{3}{2}L_\mu^2\right) \right. \nonumber \\ & 
\qquad\left.+\epsilon^2\left(-L_{\mu}(8+\frac{\pi^2}{4}) -2L_{\mu}^2-\frac{L_{\mu}^3}{2} -48-\frac{\pi^2}{3}+\zeta_3 \right) \right] \nonumber \\
& + \left(\frac{\alpha_s}{4\pi}\right)^2 C_F\left\{C_A \left(\frac{11}{2 \epsilon^2}-\frac{127}{12 \epsilon }-\frac{1705}{24}-\frac{215 L_{\mu }}{6}+5 \pi ^2-\frac{11 L_{\mu }^2}{2}-8 \pi ^2 \ln(2) + 12 \zeta _3\right) \right.\nonumber \\
& \left. +C_F \left[\frac{9}{2 \epsilon^2}+\frac{1}{\epsilon}\left(\frac{51}{4 }+9 L_{\mu }\right)+\frac{433}{8}+\frac{51 L_{\mu }}{2}-\frac{49 \pi ^2}{4}+9 L_{\mu }^2+16 \pi ^2 \log (2)-24\zeta _3\right]\right.\nonumber \\
& \left.+n_h\left[\frac{1}{2\epsilon}\left(1 + 4 L_{\mu}\right)+\frac{947}{36}+\frac{11 L_{\mu }}{3}-\frac{5 \pi ^2}{2}+3 L_{\mu }^2\right]\right.\nonumber \\
& \left.+n_l \left(-\frac{1}{\epsilon ^2}+\frac{11}{6 \epsilon }+\frac{113}{12}+\frac{19 L_{\mu}}{3}+\frac{2 \pi ^2}{3}+L_{\mu }^2\right)\right\}+ \mathcal{O}(\alpha_s^3) \, .
\end{align}

\bibliographystyle{JHEP}
\bibliography{references_inspire}

\end{document}